\documentclass[a4paper,fleqn,usenatbib]{mnras}

\usepackage[T1]{fontenc}
\usepackage{ae,aecompl}

\usepackage{subfigure}
\usepackage{graphbox,graphicx}	% Including figure files
\usepackage{amsmath}	% Advanced maths commands
\usepackage{amssymb}	% Extra maths symbols
\usepackage{color}
\usepackage{mathptmx}
\usepackage{epsfig}
\usepackage{wasysym}
\usepackage{xfrac}
\usepackage{pifont}
\usepackage[table,xcdraw]{xcolor}
\usepackage{booktabs}
\usepackage{multirow}
\usepackage{comment}
\usepackage{soul}
\usepackage{tikz}
\usepackage{orcidlink}

\definecolor{grey}{rgb}{0.75,0.75,0.75}
\definecolor{Orange}{rgb}{1.0,0.5,0.15}
\definecolor{brown}{rgb}{0.7,0.25,0.0}
\definecolor{pink}{rgb}{1.0,0.5,0.5}
\definecolor{darkerred}{rgb}{0,0.5,0.5}
\definecolor{darkerblue}{rgb}{0,0,0.8}
\definecolor{lightblue}{rgb}{0.12, 0.56, 1.0}
\definecolor{Blue}{rgb}{0,0.08,0.65}
\definecolor{Red}{rgb}{0.65,0.08,0.05}
\definecolor{Green}{rgb}{0.15,0.45,0.25}

\definecolor{purple}{rgb}{0.5,0,0.87}

\usetikzlibrary{positioning,decorations.pathreplacing,shapes}

\title[Satellite stripping efficiencies]{Stellar stripping efficiencies of satellites in numerical simulations: the effect of resolution, satellite properties and numerical disruption}

\author[G. Martin et al.]{G. Martin \orcidlink{0000-0003-2939-8668},$^{1}$\thanks{E-mail: garreth.martin@nottingham.ac.uk} F. R. Pearce \orcidlink{0000-0002-2383-9250},$^{1}$ N. A. Hatch \orcidlink{0000-0001-5600-0534},$^{1}$ A. Contreras-Santos \orcidlink{0000-0002-3374-4626},$^{2}$ A. Knebe \orcidlink{0000-0003-4066-8307},$^{2,3,4}$\newauthor 
and W. Cui \orcidlink{0000-0002-2113-4863}$^{2,3}$
\\
$^{1}$School of Physics \& Astronomy, University of Nottingham, University Park, Nottingham NG7 2RD, UK\\
$^{2}$Departamento de Física Te\'orica, M\'odulo 15, Facultad de Ciencias, Universidad Aut\'onoma de Madrid, 28049 Madrid, Spain\\
$^{3}$Centro de Investigacin Avanzada en F\'isica Fundamental (CIAFF), Facultad de Ciencias, Universidad Aut\'onoma de Madrid, 28049 Madrid, Spain\\
$^{4}$International Centre for Radio Astronomy Research, University of Western Australia, 35 Stirling Highway, Crawley, Western Australia 6009, Australia
}

\pubyear{2024}
\begin{document}
\label{firstpage}
\pagerange{\pageref{firstpage}--\pageref{lastpage}}
\maketitle


\begin{abstract}
The stellar stripping of satellites in cluster haloes is understood to play an important role in the production of intracluster light. Increasingly, cosmological simulations have been utilised to investigate its origin and assembly. However, such simulations typically model individual galaxies at relatively coarse resolutions, raising concerns about their accuracy. Although there is a growing literature on the importance of numerical resolution for the accurate recovery of the mass loss rates of dark matter (DM) haloes, there has been no comparable investigation into the numerical resolution required to accurately recover stellar mass loss rates in galaxy clusters. Using N-body simulations of satellite galaxies orbiting in a cluster halo represented by a static external potential, we conduct a set of convergence tests in order to explore the role of numerical resolution and force softening length on stellar stripping efficiency. We consider a number of orbital configurations, satellite masses and satellite morphologies. We find that stellar mass resolution is of minor importance relative to DM resolution. Resolving the central regions of satellite DM halos is critical to accurately recover stellar mass loss rates. Poorly resolved DM haloes develop cored inner profiles and, if this core is of comparable size to the stellar component of the satellite galaxy, this leads to significant over-stripping. To prevent this, relatively high DM mass resolutions of around $M_{\rm DM}\sim10^{6}$~M$_{\odot}$, better than those achieved by many contemporary cosmological simulations, are necessary.

\end{abstract}

%%%%%%%%%%%%%%%%%%%%%%%%%%%%%%%%%%%%%%%%%%%%%%%%%%

%%%%%%%%%%%%%%%%% BODY OF PAPER %%%%%%%%%%%%%%%%%%

\section{Introduction}
\label{sec:introduction}

Galaxy clusters, as the most massive collapsed structures in the Universe, serve as substantial repositories of matter and represent evolutionary endpoints in the standard $\Lambda$CDM-based hierarchical structure formation paradigm \citep[][]{Press1974,Fall1980,Moster2013}. As a consequence of their hierarchical formation, their mass assembly is expected to be dominated by the accumulation of smaller structures, encompassing a variety of scales ranging from individual galaxies to galaxy groups and other clusters. The precise accretion history of a cluster is therefore intimately tied to its cosmic environment and the specific configuration of the surrounding large-scale structure, which is responsible for funnelling matter towards the cluster. Consequently, galaxy clusters are expected to exhibit significant diversity of formation histories, which ultimately define their present-day properties.

A key luminous tracer of cluster formation history is the intracluster light \citep[ICL, see][for recent reviews]{Mihos2016,Contini2021,Montes2022}. Comprising a faint and diffuse collection of stars, the ICL is challenging to observe due to its extremely low surface brightness \citep[often below 30 mag arcsec$^{-2}$;][]{Johnston2008}. Despite this, the ICL is estimated to contribute significantly to the total stellar mass of clusters. 

Clusters in the nearby and intermediate redshift Universe have been well studied \citep[e.g.][]{Gomez1994,Feldmeier2002,Mihos2005,Adami2005,Zibetti2005,Montes2018,Kluge2020}, establishing that ICL is a ubiquitous component of galaxy clusters, comprising 10 to 40 per cent of their total stellar mass \citep{Zibetti2005,Gonzalez2007,Burke2015,Montes2018,Zhang2019,Yoo2021,Montes2022,Marx2024}. Although different definitions and methodologies can yield varying results \citep{Rudick2011,Cui2014,Brough2024}, these figures broadly align with simulation measurements \citep[e.g.][]{Rudick2006,Pillepich2014,Martin2022,Contreras2024} and semi-analytic predictions \citep[e.g.][]{Guo2011,Contini2014}. However, using these observations to infer the dynamical history of clusters or understand the processes leading to the creation of ICL remains challenging because of the complex relationship between the properties of the ICL and its formation history.

The ICL is expected to form through a variety of channels. These include (i) the pre-processing of stripped stars within galaxy groups, which are then accreted onto the cluster \citep{Mihos2004,Mihos2017,Ragusa2023}, (ii) debris from interactions between galaxies \citep{Moore1996} or (iii) as a result of stars liberated during the process of stellar stripping or disruption due to the intense tidal forces generated by the interaction between accreted objects and the cluster gravitational potential \citep[][]{Byrd1990,Gnedin2003,Purcell2007,Tang2018}. While each of these processes are thought to lead to the removal of significant quantities of baryonic and dark matter (DM) from their parent galaxies, stellar stripping is generally understood to be the dominant channel, with mergers expected to contribute a few tens of percent \citep[this can be sensitive to how the BCG or mergers are defined e.g. see][]{Murante2007,Contini2014,Contini2018}, with pre-processing becoming increasingly important in the most massive clusters \citep{Contini2014,Mihos2017,Chun2024}.

The formation of the ICL in a majority of clusters is therefore likely to proceed relatively gradually over a cluster's lifetime, with a large number of individual objects contributing different amounts of material depending on their individual physical and orbital characteristics. The properties of the ICL therefore reflect the accretion history of the cluster and the dynamical and chemical distribution of the galaxies progenerating it \citep[][]{Morishita2017,Montes2018,Contini2019,Chun2023,Chun2024}.

A robust theoretical understanding is necessary to interpret the observations and uncover a more complete understanding of the dynamical history and evolutionary state of individual clusters. In particular, addressing how the ICL is built up, as well as which objects contribute to this process. In doing so, we can connect the properties of the ICL, such as its bulk quantity \citep{Jimenez2018}, structure \citep{Mihos2004}, chemodynamical properties \citep{Sommer2005,DeMaio2018}, spatial distribution \citep{Yoo2024}, and kinematics \citep{Arnaboldi2004} to the evolutionary history of the host cluster. In light of this need, significant theoretical efforts have been made to study the formation and evolution of ICL using a variety of methods.

In addition to idealised N-body simulations \citep[e.g.][]{Johnston1999,Penarrubia2008,Errani2021} and semianalytic models \citep[e.g.][]{Napolitano2003,Watson2012,Contini2014,Tollet2017}, cosmological hydrodynamical simulations are increasingly used to study the formation of ICL. Such simulations typically achieve stellar mass resolutions of $10^{5}$~M$_{\odot}$ to $10^{7}$~M$_{\odot}$ \citep[e.g.][]{Murante2004,Willman2004,Cooper2015,Pillepich2018,Contreras2024} and have emerged as valuable tools for investigating the origin and buildup of ICL. These simulations have the advantage of self-consistently modelling the formation of clusters and their constituent galaxies within a realistic cosmological context, meaning clusters form and evolve with a realistic distribution of dynamical histories. Additionally, the formation of ICL is modelled explicitly as in N-body approaches with the addition of hydrodynamical processes such as ram pressure stripping, which observational and theoretical evidence suggest also play a minor role in the buildup of ICL \citep[][]{Tonnesen2012,George2018,Gullieuszik2020}. 

Nevertheless, the considerable computational expense of these simulations typically restricts them to coarser resolutions compared to those feasible with other approaches, introducing potential biases or uncertainties. For instance, limited mass or spatial resolution may result in the spurious disruption of bound structures due to excessively large force softening or small numbers of particles compared with those typically achieved by idealised N-body simulations \citep[][]{vandenBosch2018, Green2021}. Additionally, galaxies are only resolved down to a certain mass limit defined by the finite mass resolution of the simulation, meaning the galaxy stellar mass function becomes incomplete at lower masses. This may introduce further biases, particularly as the stars of less massive galaxies are expected to be stripped more efficiently.

Achieving and testing resolution convergence in large simulations is challenging because the dominant physical processes driving galaxy evolution, such as star formation, feedback, and cooling must be modelled using sub-grid recipes. These recipes interface at different scales depending on the numerical resolution of the simulation. Given this resolution sensitivity, resolution tests are typically restricted to studying the convergence of galaxy properties on a population level.

The success of these models in reproducing observed galaxy distributions has inspired confidence in their predictive power, with certain integrated or statistical properties of galaxies consisting of between 100 and 1000 particles often treated as reliable. However, as the convergence of specific ab-initio modelled physical processes responsible for governing the formation of ICL (such as stellar stripping) is not tested for, it is crucial to evaluate their accuracy. Establishing the reliability of these simulations in this regard is essential for understanding the potential limitations and uncertainties of this approach.

Additionally, even if bulk quantities are converged, the morphology and resolved properties of the ICL are much more sensitive to the distribution and chemodynamical properties of objects across a wide range of masses. Understanding whether stripping in lower mass galaxies -- whose total contribution to the global ICL mass budget may be small \citep[e.g.][]{Contini2014,Montes2021} but nevertheless important in shaping the ICL's resolved properties \citep[e.g.][]{Rudick2010,Kluge2024} -- is accurately resolved is vital for making valid predictions that can be benchmarked against state-of-the-art observational data.

In this paper we investigate the stellar (and DM) stripping efficiency of model satellite galaxies in a static potential. We consider satellite galaxies with a range of stellar and DM masses with either spheroidal or disc-dominated morphologies across a range of orbital configurations and simulated with a range of mass resolutions and force softening lengths in order to understand how accurately stellar stripping processes are resolved under a range of conditions.

In Section \ref{sec:method}, we describe our suite of simulations including the simulation code, galaxy models and orbital configurations used, as well as our approach to analysing the simulations. In Section \ref{sec:results}, we present our results, showing how the stripping efficiency and the resolved properties of material stripped from satellites vary as a function of the properties of the satellite, its orbit and the fidelity of the simulation. We investigate the performance of simulations across a range of resolutions typical for state-of-the-art cosmological simulations at recovering the bulk quantity of ICL and investigate how resolution influences the regions from which the stars of satellite galaxies are stripped.

\section{Method}
\label{sec:method}

\subsection{Simulation setup}
\label{sec:sims}

Our simulation setup consists of a range of satellite galaxies composed of a DM halo, with a stellar bulge and/or disc component and a cluster modelled as spherically symmetric potential.  We run a suite of N-body simulations which follow the evolution of each satellite galaxy within a static cluster potential corresponding to a $10^{14.5}$M$_{\odot}$ halo, varying the numerical and spatial resolution of the satellites.

\subsubsection{\textsc{swift}}

We make use of \textsc{swift}, an open-source gravity and smoothed particle hydrodynamics solver\footnote{http://www.swiftsim.com}. Relevant to this study, the gravity scheme implemented in \textsc{swift} employs a fast multipole method solver \citep{Greengard1987,Cheng1999} with a fixed opening angle criterion ($\theta_{cr}$) to evaluate forces between nearby particles. At larger scales, this is coupled with a standard particle mesh scheme. Gravitational softening is implemented via a \citet{Wendland1995} spline kernel. \textsc{swift} also includes support for static gravitational potentials, which we use to represent the cluster DM halo.

A more thorough description of the gravity solver in addition to the hydrodynamics solver, cosmological integration, parallelisation scheme and wide range of subgrid models for galaxy formation implemented in \textsc{swift} can be found in \citet{Schaller2023}.

Motivated by \citet{vandenBosch2018}, who have shown the rate of stripping of DM haloes is extremely robust across a very wide range of $\theta_{cr}$, we adopt a fixed opening angle of $\theta_{cr}=0.7$ for all runs in this study.

\subsubsection{Cluster potential}

We run \textsc{swift} with a static, spherically symmetric external potential representing the cluster halo. The potential is defined by a \citet[][NFW]{Navarro1996} profile (Equation \ref{eqn:NFW}), characterised by a halo mass ($M_{\rm halo}$) and scale radius ($R_{s}$), where $R_{s}$ is related to the halo concentration parameter ($c$) by $R_{s} = R_{200,c}/c$:

\begin{equation}
\label{eqn:NFW}
    \rho(r) = \frac{M_{\rm halo}}{4\pi R_s^3} \frac{1}{\frac{r}{R_s}\left(1 + \frac{r}{R_s}\right)^2}.
\end{equation}

\noindent We define a halo with a total mass of $M_{\rm halo}=10^{14.5}$~M$_{\odot}$ and a concentration parameter of $c=8$. The halo concentration is chosen primarily to maximise the dynamic range of our results while remaining consistent with the range of expected values found in the literature \citep[e.g.][]{Dutton2014b}. Our choice of $c=8$ falls within the typical range, albeit between 1$\sigma$ and $2\sigma$ above the mean concentration for a halo of this mass.

In common with other studies \citep[e.g.][]{Contini2023}, we find that larger concentrations do produce more efficient stellar stripping \footnote{Choices of $c=2, 4, 6$ and 10 yield average stripping efficiencies 0.4, 0.6, 0.8 and 1.2 times as efficient compared to our choice of $c=8$ when we keep the orbital parameters of the satellites fixed.}, however, we do not anticipate that the concentration of the host halo will significantly influence our main results, which focus on the accuracy of stripping efficiency recovery. Moreover, varying the halo concentration has a similar effect to altering the satellite's orbital pericentre, a factor we do explore in detail in this paper.

It is worth noting that real cluster haloes are not smooth or static, but are in reality dynamic systems whose properties evolve over time. This simple model therefore underestimates the stripping rate, which can be exacerbated by clumpy tidal fields \citep[e.g.][]{Knebe2006,Delos2019}, a smoothly growing potential \citep[e.g.][]{Ogiya2021} or the presence of a baryonic component in the central potential \citep{Stucker2023} among other processes. Additionally, as discussed later in Section \ref{sec:method:orbits}, this approach does not take into account dynamical friction. We do not expect ignoring any of these factors to significantly impact our main results. While factors such as satellite orbit or the composition of the central galaxy may impact stripping rates they do not significantly influence numerical accuracy.

\subsubsection{Galaxy models}

We produce model satellite galaxies by using \textsc{GalIC} \citep[][]{Yurin2014} to  specify initial conditions for an N-body system comprising a stellar bulge and disc embedded in a DM halo. \textsc{GalIC} works by iteratively optimising the velocities of particles to reach an approximate collisionless equilibrium.

Models are produced spanning total stellar masses ($M_{\star}$) from $10^{7}$ to $10^{11}$M$_{\odot}$, with halo masses drawn from the average stellar-to-halo mass relation of \citet{Moster2013}. DM haloes are modelled according to a \citet{Hernquist1990} profile and we follow \citet{Bullock2001} in assigning a spin parameter of $\lambda=0.035$ to the satellite halo. The satellite halo's scale length is chosen so that the inner region of the density profile matches an NFW halo with a concentration of $c=10$, which \citet[e.g.][]{Dutton2014b} find is the average concentration for the halo of a galaxy in the middle of our satellite halo mass range ($\sim10^{11}$~M$_{\odot}$).

We do not explore the effect of varying the mass, shape or concentration of the satellite DM halo on stripping efficiency. However, it is worth noting that the cuspiness of the inner profile has been shown to reduce stripping efficiency and promote the survival of satellite galaxies \citep[e.g.][]{Penarrubia2010} and that the haloes of most intermediate/high-mass galaxies and some low-mass galaxies are expected to exhibit cored inner profiles for at least part of their lifetime \citep{Jackson2024}.

We also do not include a gas component in our models. low-mass galaxies typically host large quantities of cold gas comparable to their total stellar mass, and the presence of this gas will act to deepen the central potential of the satellite. However, in cluster environments, gas is expected to be rapidly removed from low-mass galaxies by ram-pressure stripping, particularly for those on radial orbits \citep{McCarthy2008,Arthur2019,Kulier2023}.

Two different galaxy models are adopted for each stellar mass. The first model we consider is a spheroid-only model, where a spherically symmetric \citet{Hernquist1990} bulge component represents the entire stellar mass with its scale length set to one-tenth that of the halo scale length ($R_{s, \star} = R_{s, \rm DM}/10$)

The second model is a disc-dominated model where the disc and bulge components account for 4/5 and 1/5 of the total stellar mass respectively. The disc component is characterised by an exponential distribution whose scale length is determined by the disc's angular momentum, which is itself a fraction of the total halo angular momentum equal to the disc's mass fraction. In the vertical direction, the disc follows a \citet{Spitzer1942} sech$^{2}$-profile \citep{Springel1999}, with a constant scale height equal to 1/5 of the scale length.

We note that, in practice, cosmological hydrodynamical simulations are likely to suppress the formation of discs close to the resolution limit due to insufficient resolution, two-body heating and various other numerical issues \citep[e.g.][]{Mayer2008}. When considering poorly resolved satellites, results from our spheroidal satellite models are likely more reflective of the true impact of resolution in cosmological simulations.

Table \ref{tab:models} details the 5 spheroid and 5 disc-dominated satellite models used in this study. In order to test the effect of galaxy size on stripping efficiency, we additionally consider a set of spheroid dominated models where the size of the stellar component is varied so that $R_{s, \star} = [R_{s, \rm DM}/5, R_{s, \rm DM}/10, R_{s, \rm DM}/20, R_{s, \rm DM}/40]$, where  $R_{s, \star}  = R_{s, \rm DM}/10$ is our fiducial choice.

\begin{table}
    \centering
    \caption{Mass of the bulge, disc and halo component of each satellite model.}
    \begin{tabular}{@{}llll@{}}
        \toprule
        & $M_{bulge}/10^{10}~\rm{M_{\odot}}$ & $M_{disc} /10^{10}~\rm{M_{\odot}}$ & $M_{halo} /10^{10}~\rm{M_{\odot}}$ \\ \midrule
        \textbf{Spheroid} & 0.001                         & 0                             & 1.8                           \\
        & 0.01                          & 0                             & 4.2                           \\
        & 0.1                           & 0                             & 13.2                           \\
        & 1                             & 0                             & 39.1                          \\
        & 10                            & 0                             & 152.5                          \\
        \midrule
        \textbf{Disc-} & 0.0002                        & 0.0008                        & 1.8                           \\
        \textbf{dominated} & 0.002                         & 0.008                         & 4.2                           \\
        & 0.02                          & 0.08                          & 13.2                           \\
        & 0.2                           & 0.8                           & 39.1                          \\
        & 2                             & 8                             & 152.5                          \\ \bottomrule
    \end{tabular}
    \label{tab:models}
\end{table}

\subsubsection{Resolution}

To investigate convergence in a scenario close to typical studies using cosmological simulations, we run a set of `low DM resolution' simulations, intended to approximately correspond to mass and spatial resolutions achieved by contemporary cosmological hydrodynamical simulations \citep[e.g.][]{Dubois2014,Hirschmann2014,Schaye2015,Bahe2017,Nelson2019}. We pick stellar particle mass resolutions of $10^{5}~$M$_{\odot}$, $10^{6}~$M$_{\odot}$ and $10^{7}~$M$_{\odot}$ and softening lengths of 0.3~kpc, 0.7~kpc and 1.4~kpc. For the corresponding DM mass resolution, we adopt mass resolutions of $10^{6}~$M$_{\odot}$, $10^{7}~$M$_{\odot}$ and $10^{8}~$M$_{\odot}$, 10 times larger than the stellar mass resolution, and adopt softening lengths identical to those used for the stars.

The DM halo is expected to delay stripping of the stellar component in satellites, at least initially \citep[e.g][]{Pfeffer2013,Contini2017}. DM haloes in clusters undergo considerably more rapid stripping than the stellar component, losing approximately 80 percent of their infall mass and reducing the extent of the DM component close to that of the stellar radius after just one or two pericentric passages, \citep[e.g.][]{Villalobos2012,Smith2016,Joshi2019,Haggar2021}. However, DM is still expected to remain dominant over stars, and the DM halo is therefore expected to continue to strongly influence stellar stripping efficiencies even after a large fraction of DM is liberated.

We therefore conduct a set of `high DM resolution' runs, where the stellar mass resolution matches that of the low DM resolution runs, but the halo is consistently resolved with a large number of particles. Our choice of numerical resolution for the DM is motivated by \citet{Errani2021} and \citet{Benson2022}, who show that the density profiles of tidally stripped subhaloes begin to show systematic deviations from their predicted evolutionary `tidal tracks' \citep[e.g.][]{Penarrubia2008} once they are stripped to fewer than a few thousand particles. Contemporary large-scale cosmological simulations resolve the DM haloes of low-mass galaxies with numerical resolutions significantly below this benchmark, potentially resulting in more rapid stripping of the DM halo due to artificial disruption \citep{vandenBosch2018, Errani2021}. 

We choose an initial particle resolution of $N_{\rm DM}=2^{17}$, which we find is sufficient to ensure that all haloes remain resolved above the limit of a few thousand particles found by \citet{Errani2021} for the full duration of our simulations. The choice of force softening length is complex, as a softening length that is optimal for an isolated galaxy will not remain optimal in a dynamic environment, where the size or concentration, and mass of the remnant halo varies with the tidal evolution of the satellite \citep[e.g.][]{vanKampen2000}. Following \citet{vanKampen2000}, we choose a softening length based on the mean interparticle separation of stars within the half-mass radius of the initial conditions of the satellite. Between our least and most massive satellites, we find optimal softening between 0.25~kpc and 1.15~kpc. We opt for a common softening length of 1~kpc for all satellites, which lies between these values, and is comparable to softening lengths used in our low DM resolution runs. 

Finally, we run a series of `benchmark' simulations using the same DM resolution as the high DM resolution run while also resolving the stellar component of every satellite with $2^{17}$ star particles. As with the DM, we choose an optimum stellar force softening length based on the mean interparticle separation of stars within the half-mass radius of the initial conditions of the satellite. For the $2^{17}$ star particle model, this corresponds to a softening length between 0.025~kpc for the lowest mass satellite and 0.1~kpc for the highest mass satellite. We opt for a common softening length of 0.05~kpc for all satellites, roughly in the middle of these two values. We also keep the softening length fixed at 0.05~kpc for all lower resolution runs. As we show later in Sections \ref{sec:softening_stripping} and \ref{sec:fixed_n}, our results are not very sensitive to the softening length chosen for the stars, but depend much more significantly on the properties of the satellite DM halo.

For each of the runs described above, we also run multiple simulations for the most poorly resolved satellites at each resolution level (those resolved with only 100 star particles). For these satellites, we produce 10 different realisations of the initial conditions. All quantities presented in this paper for these poorly resolved satellites are given as an average of these 10 realisations.

Table \ref{tab:resolution} summarises the resolution and softening lengths used for each set of runs.

\begin{table}
    \centering
    \caption{Numerical resolution and force softening lengths ($\epsilon$) used for the stellar and DM components of the satellite models in different runs. Numerical resolution is indicated as either the particle mass ($m_{\star}$, $m_{\rm DM}$) or number of particles ($N_{\star}$ or $N_{\rm DM}$). Our benchmark run is denoted in bold.}
    \begin{tabular}{@{}lllll@{}}
    \toprule
    High DM resolution & $m_{\star}/\rm{M_{\odot}}$ & $\epsilon_{\star}/\rm{kpc}$ & $N_{\rm DM}$ & $\epsilon_{\rm DM}/\rm{kpc}$ \\  
    \midrule
    \multirow{3}{*}{}     & $10^{5}$ & 0.3 & $2^{17}$ & 1.0           \\
                          & $10^{6}$ & 0.7 & $2^{17}$ & 1.0            \\
                          & $10^{7}$ & 1.4 & $2^{17}$ & 1.0            \\ 
    \midrule

    Low DM resolution & $m_{\star}/\rm{M_{\odot}}$ & $\epsilon_{\star}/\rm{kpc}$ & $m_{\rm DM}/\rm{M_{\odot}}$ & $\epsilon_{\rm DM}/\rm{kpc}$ \\  
    \midrule
    \multirow{3}{*}{}     & $10^{5}$ & 0.3 & $10^{6}$ & 0.3           \\
                          & $10^{6}$ & 0.7 & $10^{7}$ & 0.7           \\
                          & $10^{7}$ & 1.4 & $10^{8}$ & 1.4            \\ 
    \midrule

    Fixed particle number & $N_{\star}$  & $\epsilon_{\star}/\rm{kpc}$ & $N_{\rm DM}$ & $\epsilon_{\rm DM}/\rm{kpc}$  \\ 
    \midrule
    \multirow{4}{*}{} \textbf{Benchmark}    & $\mathbf{2^{17}}$           & \textbf{0.05} & $\mathbf{2^{17}}$ & \textbf{1.0}             \\
%                          & $2^{14}$           & 0.05 & $2^{17}$ & 1.0            \\
%                          & $2^{11}$           & 0.05 & $2^{17}$ & 1.0            \\
%                          & $2^{8}$            & 0.05 & $2^{17}$ & 1.0            \\ 
    \bottomrule
    \end{tabular}
    \label{tab:resolution}
\end{table}

As a final note on force softening lengths, we would like to point out that we employ relatively large fixed softening lengths in this work. These softening lengths are consistent with those commonly used in cosmological simulations. Although reducing force softening lengths can help alleviate issues such as the spurious disruption of subhaloes \citep{Hopkins2023}, this approach is not practical for cosmological simulations. This is because mitigating spurious halo disruption in clusters must be balanced against the impact of softening lengths on the artificial suppression or fragmentation of haloes during structure formation \citep[e.g.][]{vanKampen2000,Iannuzzi2011,Mansfield2021}, which can significantly affect initial halo abundances. 

Some available astrophysical codes implement N-body gravity solvers with adaptive force softening \citep[e.g.][]{Teyssier2002,Price2007,Hopkins2023}. Such schemes may help mitigate some issues such as artificial disruption of substructures and spurious two-body scattering,  which impact stellar stripping rates. Consequently, some of the issues identified in this study may be less pronounced in simulations using these adaptive codes.

\begin{figure*}
    \centering
    \includegraphics[width=0.95\textwidth]{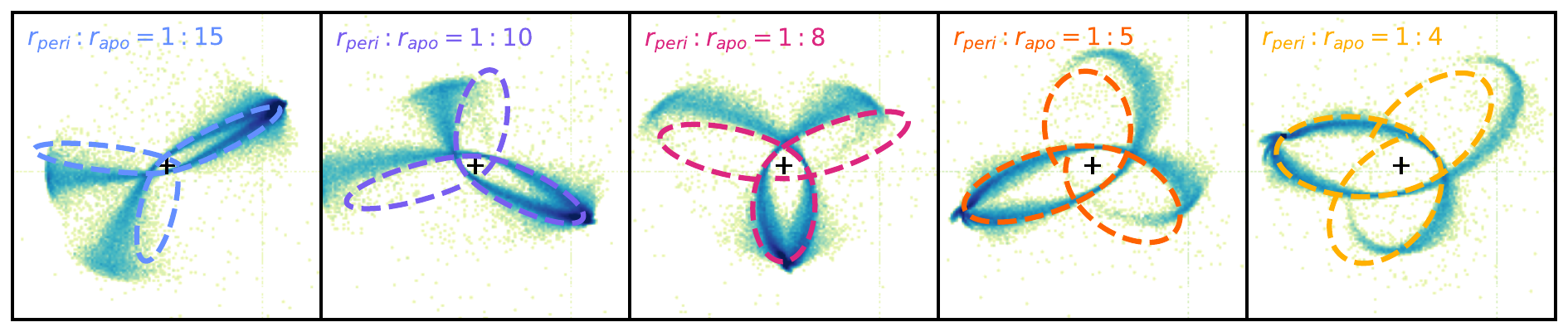}
    \caption{Each panel shows the stellar debris and surviving satellite for the $M_{\star}=10^{8}~{\rm M_{\odot}}$ spheroid model after $t=7$~orbits for a different orbital configuration. 
    Points indicate the distribution and are coloured according to the local surface density of star particles with a logarithmic colour scale. Dashed lines indicate the path of the satellite over the previous three orbital periods and the centre of the potential is marked by a black cross.}
    \label{fig:orbits}
\end{figure*}

\subsubsection{Orbits}
\label{sec:method:orbits}

%and satellites whose pericentres are closest to the central galaxy are expected to be dominated by highly radial orbits
Finally, we choose a range of orbital configurations with pericentre to apocentre ratios ($r_{\rm peri}:r_{\rm apo}$) ranging from 1:4 to 1:15 where the apocentre remains fixed at 1500 kpc, as detailed in Table \ref{tab:orbits}. In general, highly circular orbits are expected to be very rare regardless of mass ratio, with most infalling satellites travelling on orbits taking them close to the centre of the host halo. For example, despite considering very different mass ratios (major galaxy mergers compared with satellites of cluster mass haloes), both \citet{Khochfar2006} and \citet{Wetzel2011} show that around 85 per cent of galaxies have orbits with $r_{\rm peri}:r_{\rm apo}$ ratios smaller than 1:4. Similarly, \citet{Smith2015} show a majority of low-mass satellites travel on highly eccentric orbits within a Virgo-like halo. We therefore do not consider more circular orbits which, as previous work \citep[e.g.][]{Smith2015} and the results of Section \ref{sec:results} show, are in any case subject to slow rates of stellar stripping and therefore contribute little material to the ICL.

Satellites are injected at the apocentre of their respective orbits and the simulation is run for 10 orbital periods or around 5~Gyrs. This corresponds roughly to the formation time of a halo that reaches a mass of $10^{14.5}$~M$_{\odot}$ by $z=0$ \citep[][]{McBride2009} . Figure \ref{fig:orbits} shows an example of each orbit from Table \ref{tab:orbits} along with maps showing the surface density of the stellar debris.

As we only model the cluster as an external potential, our simulations neglect the effect of dynamical friction, which would otherwise cause orbits to gradually lose energy and sink towards the centre of the potential. However, for the satellite-to-cluster halo mass ratios considered here, which never exceed 2 percent, the merging timescales due to dynamical friction are expected to exceed the duration of our simulations ($>5$~Gyrs), even for highly radial orbits \citep{Boylan2008}. As a result, mergers with the central galaxy are unlikely, even if dynamical friction were included.

For satellites more massive than those considered here, mergers are expected to begin contributing to the production of ICL. In such interactions, violent relaxation \citep{LyndenBell1967} becomes the dominant process rather than stripping. Consequently, ICL production due to mergers is likely to be subject to different resolution effects and may influence the properties of the ICL in distinct ways \citep{Moore1996,Murante2007,Kluge2020,Joo2023}, although mergers are expected to make a more minor contribution to the ICL overall \citep[][]{Contini2014,Burke2015}.

While dynamical friction is absent from our simulations, dynamical self-friction resulting from tidal mass loss \citep[][]{Fellhauer2007} does lead to a gradual decay of each satellite from its initial orbit. We confirm that no orbit experiences no more than a 3 per cent decay in its apocentric radius over the full 10 orbits, meaning satellite orbital configurations, and therefore the tidal field felt by each satellite remains essentially unchanged over the timescales considered in this study. %Although the impact of this dynamical self-friction remains modest, it becomes more pronounced with increasing satellite mass. However, we confirm that, for our most massive satellite, the orbit experiences no more than a 3 per cent decay in its apocentric radius over the full 10 orbits.

\begin{table}
\centering
\caption{Configuration for each of the orbits considered in this study. Pericentric radius ratio of the pericentric and apocentric radius, and total number of orbits. The apocentric radius is kept fixed at 1500~kpc.}
\label{tab:orbits}
\begin{tabular}{@{}lll@{}}
\toprule
$r_{\rm peri}/\rm{kpc}$ & $r_{\rm peri}:r_{\rm apo}$ & $N_{\rm orbits}$ \\ \midrule
100        & 1:15                       & 10               \\
150        & 1:10                       & 10               \\
200        & 1:8                        & 10               \\
300        & 1:5                        & 10               \\
400        & 1:4                        & 10               \\ \bottomrule
\end{tabular}
\end{table}

\subsection{Galaxy tracking}

In idealised simulations, a commonly employed method for determining which particles are bound to an object involves initially calculating the centre of mass of the particles. This can be achieved, for instance, by computing the centre of mass within a shrinking sphere \citep[e.g.][]{Power2003} or by tracking the highest density point, \citep[e.g.][]{Penarrubia2008, Chang2013}. Subsequently, particles bound to the object are identified based on this centre. However, as the reliability of determining the centre of mass diminishes when dealing with a small number of particles, we opt for an alternative approach in which we follow the 10 most bound particles and define the centre of the galaxy as their median position, defining which stars are stripped based on a fixed aperture. We determine when a star has been stripped from a satellite according to the following steps:

\begin{enumerate}
    \item \noindent Initially, we calculate the binding energy of each particle in the initial conditions (at $t=0$) by summing their kinetic and potential energy. We identify the 10 most strongly bound particles and set the centre of the galaxy to be their median position.\\
    \item \noindent Subsequently, the radii $R_{\rm bound,\star}$ and $R_{\rm bound,DM}$ are calculated, defined as 1.5 times the radius from our calculated galaxy centre containing 90 percent of the star and DM particles, respectively\footnote{We note that the stripped fractions we recover remain robust regardless of the exact values of $R_{\rm bound,\star}$ and $R_{\rm bound,DM}$, as even significant increases in $R_{\rm bound}$ have minimal impact on the final stripped fractions. For example increasing the value $R_{\rm bound,\star}$ by five times, yields a variation in the final stripped fractions of just 1 per cent.}.\\
    \item \noindent Next, the galaxy is tracked in subsequent snapshots by following the centre defined by the 10 most bound particles. At each snapshot:
    \vspace{-\topsep}
    \begin{enumerate}
        \item We first identify the centre of the object as the new median position of the 10 most bound particles identified at $t=0$.
        \item Next, we identify all DM particles within $R_{\rm bound,DM}$ and any star particles within $R_{\rm bound,*}$.
        \item Any particles that have remained outside of $R_{\rm bound}$ for a time exceeding one quarter of an orbital period are deemed to have been stripped at the snapshot where they were initially detected beyond $R_{\rm bound}$. If any of these stripped particles correspond to those identified as the most bound particles at $t=0$, they are disregarded during the determination of the object centre in subsequent snapshots.
    \end{enumerate}
\end{enumerate}

We verify that the galaxy centre has been correctly determined through comparison with the orbit of a test particle placed in the same gravitational potential. Although the test particle's orbit is not expected to perfectly mirror that of the galaxy due to the effect of dynamical self-friction, our analysis consistently shows only minor deviations between the two.

\section{Results}
\label{sec:results}

\subsection{How does DM resolution effect stellar stripping efficiency?}

\subsubsection{Numerical resolution}
\label{sec:fixed_n}

We first explore the role of satellite DM resolution on stellar stripping efficiencies.

\begin{figure}
    \centering
    \includegraphics[width=0.45\textwidth]{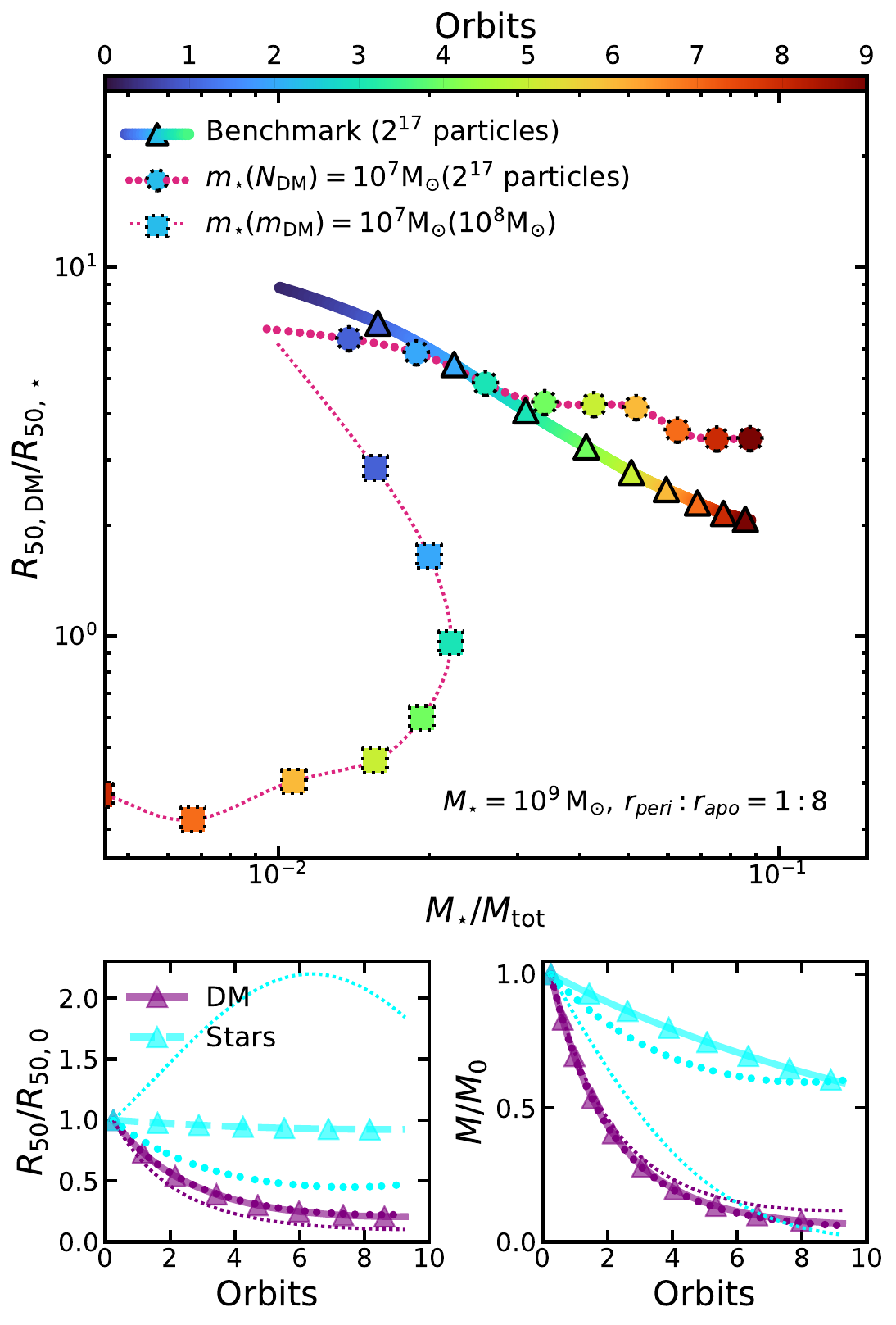}
    \caption{\textbf{Top}: Tracks showing the evolution of the ratio between DM and stellar half-mass radii ($R_{50, \rm {DM}} / R_{50, \rm {\star}}$) and the stellar mass fraction ($M_{\star}/M_{\rm tot}$) for an example $M_{\star}=10^{9}$~M$_{\odot}$ spheroidal satellite on an orbit with $r_{\rm peri}:r_{\rm apo} = 1:8$ in different resolution runs. The benchmark run is indicated with a thick multicoloured line where the colour shows the number of orbits elapsed as indicated by the colourbar at the top of the panel and triangle markers indicate whole numbers of orbits. The `high DM resolution' and `low DM resolution' runs are indicated by a tick dotted line and finely dashed line respectively, with circle and square markers indicating whole numbers of orbits. Their respective stellar and DM mass resolutions are indicated in the legend. \textbf{Bottom}: Evolution of the half-mass radius (left) and mass (right) of the DM (purple) and stellar (cyan) components of the satellite in the same benchmark, high DM resolution and low DM resolution runs.}
    \label{fig:DM_stripping}
\end{figure}

In the top panel of Figure \ref{fig:DM_stripping} we present tracks showing the time evolution of the DM half-mass radius ($R_{50, \rm {DM}}$) and stellar half-mass radius ($R_{50, \rm {\star}}$) ratio as a function of the stellar mass fraction for an example spheroidal satellite with a stellar mass of $M_{\star}=10^{9}$~M$_{\odot}$. Results are shown for the benchmark simulation (thick coloured line with up-pointing triangles) as well as example 'high DM resolution' (thick dotted line with circle markers) and 'low DM resolution' (thin dashed line with square markers) runs. To clearly demonstrate the impact of resolution we present results for the lowest resolution level for each of the runs. For both the high DM resolution and low DM resolution runs we use a stellar mass resolution of $10^{7}$~M$_{\odot}$ (corresponding to only 100 star particles). The high DM resolution run has the same DM mass resolution as the benchmark run ($2^{17}$ particles) and for the low DM resolution run we use a DM mass resolution of $10^{8}$~M$_{\odot}$ (corresponding to 1320 DM particles).

Line colour indicates the number of orbits undergone and each coloured triangle, circle and square marker indicates a full number of orbits. While we show our result for a single satellite model and orbital configuration, the results presented for the low DM resolution run are representative of the most poorly resolved satellites at any given resolution level. Additionally, we do not find any qualitative difference in our results across different orbital configurations.

Although the DM component is stripped more quickly than the stellar component, it continues to dominate in terms of mass over the entire duration of all three simulation runs (this is true of satellites of all masses considered). The properties of the stellar component are therefore dictated by those of the DM component regardless of the mass resolution.

In the benchmark and high DM resolution runs, we find that the satellites become consistently more stellar mass dominated over time, indicating that the relative stripping rate of the DM component is considerably faster than the stellar component, in common with previous studies \citep[e.g.][]{Smith2016}. As the DM halo is considerably more extended than the stellar component, more weakly bound DM particles far from the centre of the potential well are more easily stripped when the satellite's orbit passes through the dense centre of the cluster. At the same time, the DM radius quickly shrinks so that it is equal to only four or five times the stellar half-mass radius after just a few orbits in both the benchmark and high DM resolution runs.

%Despite significant differences in the initial stellar mass fraction of each satellite, we observe relatively uniform behaviour in the benchmark run tracks. We also observe only limited variation in tracks across different orbits, with more radial orbits becoming less DM dominated slightly more quickly. 

We observe significant differences in the low DM resolution run. There is initially a slower increase in the stellar mass fraction than in the benchmark and high DM resolution runs and after three or four orbits the stellar mass fraction actually begins to become smaller. At the same time, the DM radius approaches the stellar radius much more rapidly, reaching the stellar half mass radius after a similar number orbits. This means that the stellar radius reaches the stripping radius far more quickly than in the benchmark run.

The bottom two panels of Figure \ref{fig:DM_stripping} show separately the evolution of half-mass radius and remaining mass of the DM and stellar components of the same satellite relative to their $t=0$ values for the same three runs. It is clear that the differences observed between the benchmark and low DM resolution runs are not driven by differences in the stripping rate or overall size evolution of the DM, whose evolution remains similar across all three simulation runs. Instead, we observe significant differences in the properties of the stellar component, which experiences more efficient stripping and significant size growth at lower DM mass resolution. In the low DM resolution run, the stellar component doubles in size, whereas its radius shrinks by around 50 per cent and 5 per cent in the high DM resolution and benchmark case respectively.

The stellar stripping efficiency only differs significantly in the low DM resolution run, with the benchmark and high DM resolution runs resulting in similar stellar mass loss rates and the low DM resolution run resulting in far more rapid stellar stripping. The similarity between the benchmark and high DM resolution runs indicates that the DM mass resolution rather than stellar mass resolution is driving these effects. However, this must be the result of something more subtle than global properties like the total mass or size of the satellite's halo, which the bottom panels of Figure \ref{fig:DM_stripping} show do not change appreciably across the different resolution levels.

The shrinking stellar half-mass radius, which is not associated with significantly more efficient stellar stripping, in the high DM resolution run is a result of discretization noise, not the properties of the halo. Discretization noise causes more particles to be stripped from smaller radii as discussed further in Section \ref{sec:location}.

On the other hand the increased stellar half-mass radius and rapid stellar stripping are a result of the smaller scale properties of the halo. All haloes have a poorly resolved central region whose radius increases with lower numerical resolution. Within this central unresolved region, a large artificial core can be produced. We explore how resolution affects the central part of the DM halo by first defining the inner DM profile slope, measured as the power-law slope between $R_{50, \rm {\star}}$ and 2~$R_{50, \rm {\star}}$ ($\alpha_{\rm DM}[1-2~R_{50, \star}]$), where $R_{50, \star}$ is measured at $t=0$. When averaging over all orbits and satellite masses, we find that, by the second apocentric passage, the average inner slope has evolved from an expected value of $\alpha_{\rm DM}[1-2~R_{50, \star}]=-1$ to a value $\alpha_{\rm DM}[1-2~R_{50, \star}]=-0.514\pm0.008$ for the benchmark run and $\alpha_{\rm DM}[1-2~R_{50, \star}]=-0.138\pm0.019$ for the low DM resolution run, maintaining similar values over subsequent orbits in both cases.

In line with previous studies \citep[e.g.][]{Power2003, Springel2008}, we see that halo density profiles flatten significantly within the unresolved region. For the example satellite shown in Figure \ref{fig:DM_stripping}, we find an almost completely flat (i.e. cored) inner DM profile, with similar results obtained for similarly poorly resolved satellites. The stellar component becomes significantly more extended in response to the flatter central potential, resulting in considerable evolution of the half-mass radius of the stellar component. In turn, the stripping of stars lying within this flat inner core becomes significantly more efficient than would be the case in a more cuspy halo.

\subsubsection{DM and stellar force softening lengths}
\label{sec:softening_stripping}

Previous studies \citep[e.g.][]{vanKampen2000,vandenBosch2018,Mansfield2021} have emphasised the importance of choosing appropriate force softening lengths for accurately recovering the properties and stripping efficiencies of DM haloes. In this section, we extend this to test the effect of these choices on stellar stripping efficiency by studying how stellar and DM force softening length influence the stellar stripping rates at a fixed numerical resolution.

\begin{figure}
    \centering
    \includegraphics[width=0.45\textwidth]{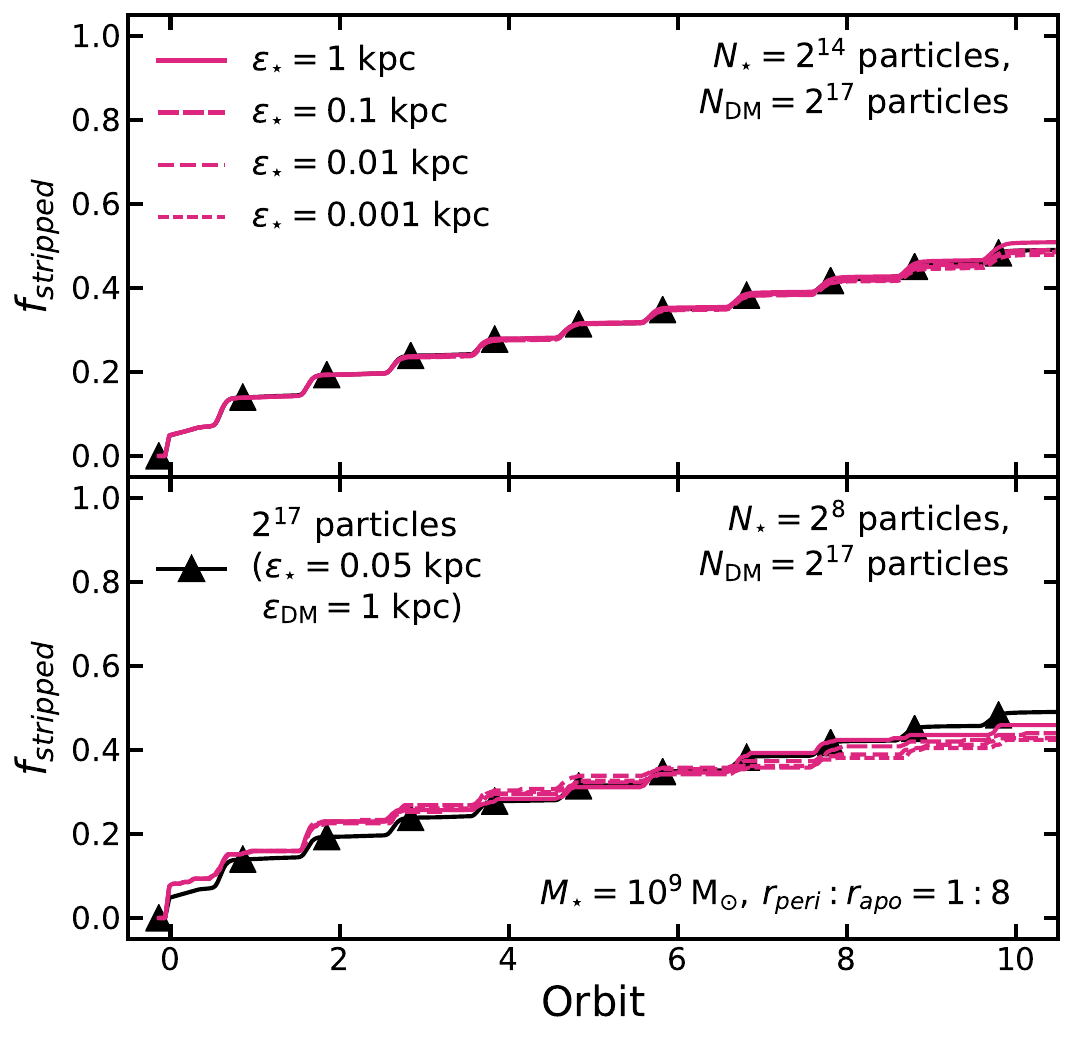}
    \caption{Tracks indicating the  mass of stars stripped as a fraction of the initial mass of a $10^{9}$~M$_{\odot}$ spheroidal satellite simulated with different stellar force softening lengths. The top and bottom panel show results for the same satellite resolved with $2^{14}$ and $2^{8}$ star particles respectively. Different softening lengths are indicated by different line styles and the benchmark run is indicated by a black line with up-pointing triangle markers.}
    \label{fig:softening_stripping}
\end{figure}

Figure \ref{fig:softening_stripping} shows the mass of stars stripped as a fraction of the initial mass for an example satellite. Stripped fractions are shown as a function of orbital period with each whole number of orbits indicating the pericentre. Different stellar softening lengths correspond to different line styles indicated in the legend and the benchmark ($2^{17}$ particle, $\epsilon_{\star}=0.05$~kpc) run is indicated by a black line marked with up-pointing triangles. The mass resolution and softening length of the DM halo are kept fixed across all runs.

The top panel shows our result for a satellite resolved with $2^{14}$ particles. We observe almost no variation in the stripping efficiency at different stellar force softening lengths. The bottom panel shows the same result for a satellite resolved with $2^{8}$ particles. Again, the level of variation is small, albeit slightly larger than for the better resolved satellite. In this example, we have shown results for a spheroidal satellite, but we see qualitatively similar results for disc-dominated satellites.

\begin{figure}
    \centering
    \includegraphics[width=0.45\textwidth]{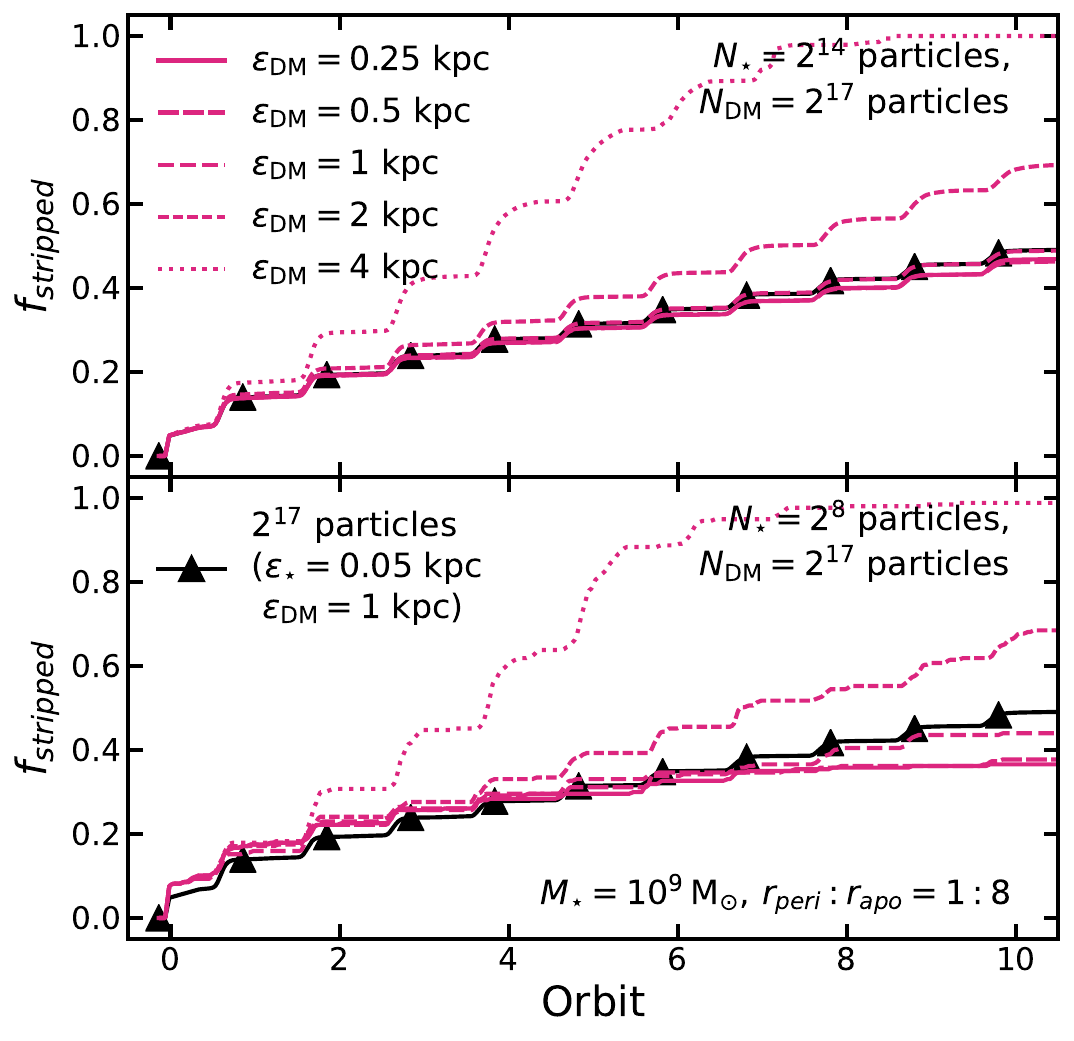}
    \caption{Tracks indicating the  mass of stars stripped as a fraction of the initial mass of a $10^{9}$~M$_{\odot}$ spheroidal satellite simulated with different DM force softening lengths. The top and bottom panel show results for the same satellite resolved with $2^{14}$ and $2^{8}$ star particles respectively. Different softening lengths are indicated by different line styles and the benchmark run is indicated by a black line with up-pointing triangle markers.}
    \label{fig:softening_stripping_DM}
\end{figure}

Figure \ref{fig:softening_stripping_DM} shows the same result as Figure \ref{fig:softening_stripping}, instead varying the DM force softening length and fixing the stellar force softening length at 0.05~kpc. In this case, there is a clear increase in stripping efficiency for force softening lengths larger than 1~kpc. This is not a direct result of the DM profile becoming physically shallower. According to the same definition as above, DM inner power-law slopes, $\alpha_{\rm DM}[1-2~R_{50, \star}]$, are consistent with those found for the benchmark run in Section \ref{sec:fixed_n}, regardless of the DM force softening length used.

Isolated simulations with the same force softening but no external potential retain slightly steeper central profiles (with logarithmic slopes steeper by between 0.1 and 0.2), indicating some modest evolution in the inner DM profile due to the tidal field rather than numerical effects. However, regardless of whether the satellite is isolated or experiences a tidal field, we find that the half-mass radius of the stellar component always expands to around the force softening length provided the initial effective radius of the galaxy exceeds the force softening length.

If an excessive DM force softening is chosen, this can therefore lead to star particles being subject to an over-softened central potential from the dominant DM component. This is distinct from expansion of the stellar component driven by insufficient sampling at the centre of the potential as shown in Section \ref{sec:fixed_n} -- it is not the result of a physical change in the central DM profile, but rather due to the stars being subject to a different force law at small separations.

Based on our findings, the stellar softening length appears to be of minor importance to stripping efficiency; although our choice of stellar softening lengths in Figure \ref{fig:softening_stripping} ranges over four orders of magnitude, we see remarkably little impact on stripping efficiency. As we have shown in Figure \ref{fig:DM_stripping}, the DM halo always remains dominant over the stellar component. The motion of the stars is therefore still largely determined by the halo and, for this reason, the stellar force softening length has very little impact compared with the DM force softening length.

The most important factor for the accurate recovery of stellar stripping efficiencies is that the centre of the potential is well resolved, as stellar stripping proceeds significantly more efficiently when the stellar component resides in a flattened potential. As discussed later in Section \ref{sec:recovering_ICL_fraction}, this may result in smaller galaxies being over-stripped compared to equivalent mass galaxies with larger sizes.

\subsection{How well converged are stripping rates for different stellar masses?}

\subsubsection{Stripping efficiency as a function of resolution and satellite properties}
\label{sec:fixed_m}

In this section, we perform a detailed study on the effect that numerical resolution has on the stripping efficiency of satellites with different masses and morphologies. We consider four different resolution levels: the benchmark run, where all satellites are resolved with $2^{17}$ DM and star particles, and the three `low DM resolution' runs, where we have picked stellar mass resolutions and softening lengths representative of those typically achieved by large cosmological simulations (see Table \ref{tab:resolution}). We also show results for the `high DM resolution' runs, where the stellar components of the satellites are resolved with the same resolution level as the low DM resolution runs, but the DM component is always resolved with $2^{17}$ particles.

\begin{figure*}
    \centering
    \includegraphics[align=t, width=1\textwidth]{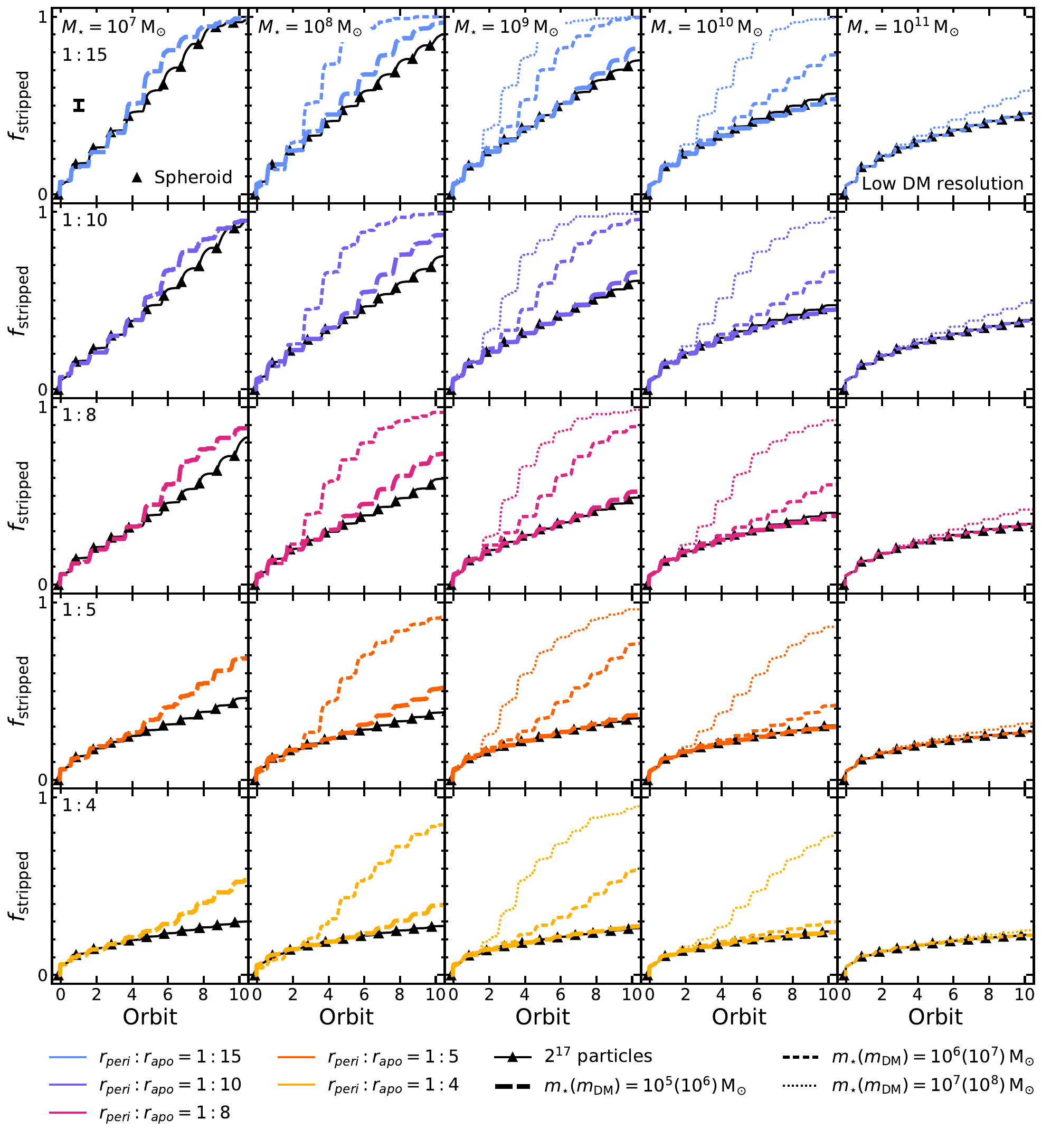}
    \caption{Tracks indicating the mass of stars stripped as a fraction of the initial mass for different mass spheroidal satellites for the low DM resolution runs. More finely dashed, thinner lines indicate lower mass resolutions and the benchmark resolution is indicated by black lines with up-pointing triangle markers. Different colour lines indicate different orbital configurations. For the lowest resolution run, tracks shown as an average of 10 realisations, with the black error bar in the top-left panel indicating the variation in stripped fractions after 10 orbits over the 10 realisations averaged over the lowest resolution satellites at all masses. For clarity, we only show results for spheroids, but an example of tracks for both disc-dominated and spheroidal satellites is shown in Figure \ref{fig:example_stripping}.}
    \label{fig:stripping_tracks_sph_I_lr}
\end{figure*}

In Figure \ref{fig:stripping_tracks_sph_I_lr}, we present tracks showing the fraction of initial stellar mass stripped since infall ($f_{\rm stripped}$) for spheroidal satellites with different stellar masses for our low DM resolution runs. From left-to-right panels correspond to satellites of increasing stellar mass and from bottom-to-top, panels indicate increasingly radial orbital configurations, which are indicated at the top-left corner of each panel. Each panel shows tracks indicating the stripped fractions as a function of orbital period with each whole number of orbits indicating the pericentre. The benchmark resolution run is indicated by a solid black line marked with up-pointing triangles and fixed stellar mass resolutions of $m_{\star}=10^{5}$~M$_{\odot}$, $10^{6}$~M$_{\odot}$ and $10^{7}$~M$_{\odot}$ are indicated as increasingly thin, and increasingly finely dashed lines.

Panels corresponding to objects with stellar masses of $10^{7}$~M$_{\odot}$ and $10^{8}$~M$_{\odot}$ do not show tracks for all resolution levels, as some objects remain entirely unresolved. Black error bars in the left of the first panel indicate the range in $f_{\rm stripped}$ after 10 orbits averaged over 10 realisations of the most poorly resolved satellites (those with 100 particles).

We observe alternate phases of fast mass loss at the pericentre of the orbit followed by a longer period of much more gradual mass loss as the satellite approaches the apocentre. By far a majority of mass loss occurs in the first phase across all satellites properties and all resolution runs, indicating that the mass evolution of the satellite remains well described by an impulse approximation \citep{Gnedin1999}, regardless of orbital configuration, properties of the satellite or resolution.

For the lowest resolution level, we observe considerable overstripping compared with the benchmark run for satellite stellar masses smaller than $10^{11}$~M$_{\odot}$ ($N_{\star}<10,000$ and $N_{\rm DM}<15,000$). Stellar stripping tracks in the highest resolution non-benchmark run remain converged down to $m_{\star}=10^{9}$~M$_{\odot}$, beginning to diverge slightly at lower satellite masses.

\begin{figure}
    \centering
    \includegraphics[width=0.45\textwidth]{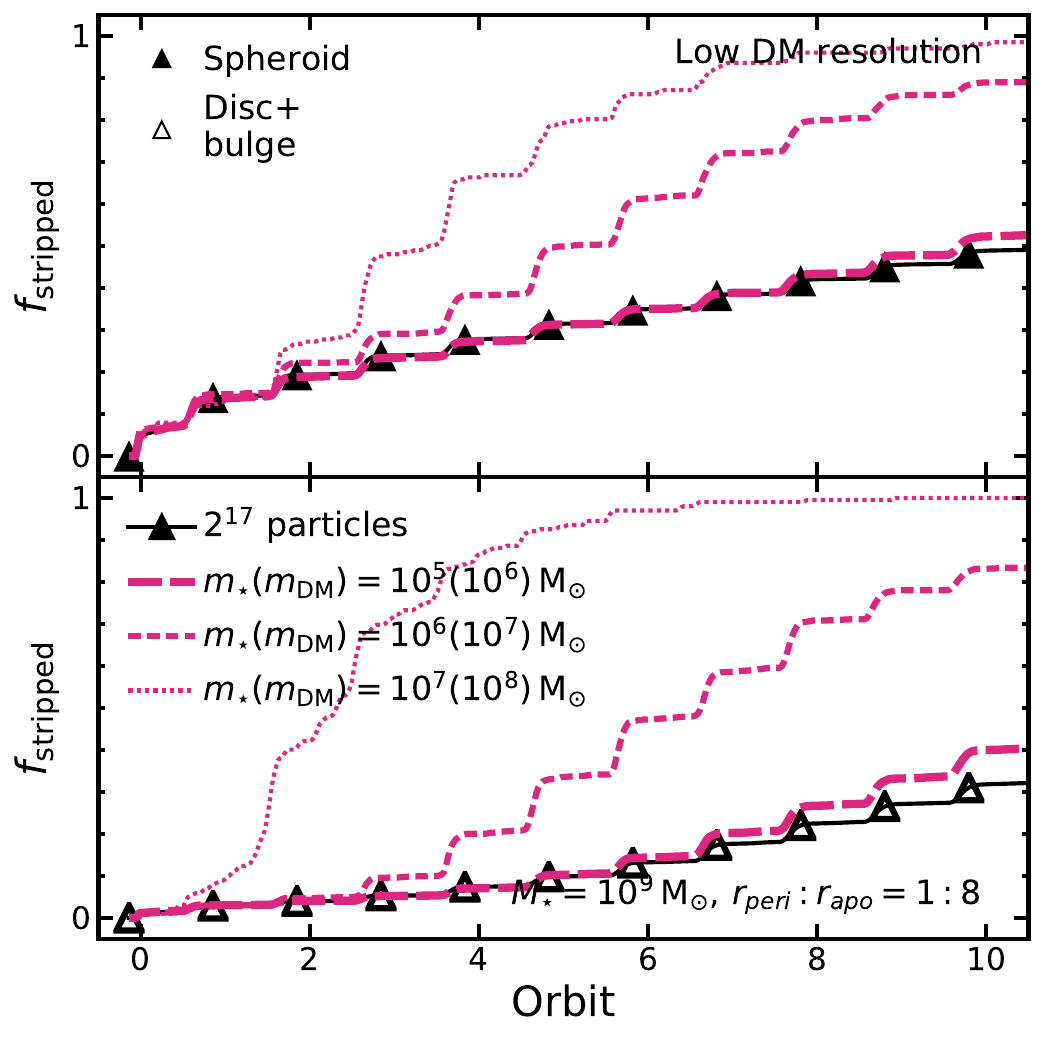}
    \caption{Tracks indicating the  mass of stars stripped as a fraction of the initial mass of a $10^{9}$~M$_{\odot}$ spheroidal (top panel) and disc-dominated (bottom panel) satellite. More finely dashed, thinner lines indicate lower mass resolutions and the benchmark run is indicated by a black line with up-pointing triangle markers.}
    \label{fig:example_stripping}
\end{figure}

In Figure \ref{fig:example_stripping} we show a comparison of the stripped fractions for disc-dominated and spheroidal satellites for a satellite with $M_{\star}=10^{9}$~M$_{\odot}$ and $r_{\rm peri}:r_{\rm apo}=1:8$. The example shown is generally representative of the trends seen for other satellite masses and orbital configurations.

In the benchmark runs, we find that the stripping efficiency of disc-dominated satellites is initially lower than for spheroidal satellites. In our disc-dominated models, a diffuse component of the bulge extends beyond the edge of the disc, meaning the least strongly bound and most extended part of the bulge is stripped before the disc begins to lose mass. Once this component of the bulge has been stripped to the radius of the disc, the disc begins to dominate. Subsequently, the stellar stripping rate increases due to the disc's higher stellar density. For satellites with the most circular orbits, where the stripping radius may never reach the radius of the disc, stripping efficiency remains low because only the diffuse outskirts of the bulge component are ever stripped. At different resolution levels, the amount of over-stripping has some dependence on morphology. Satellites typically diverge from the benchmark tracks at similar points regardless of morphology, but the level of divergence is generally larger for disc-dominated satellites.

We avoid drawing any specific conclusion regarding relative differences in stripping efficiency of real disc or spheroid dominated galaxies as our satellites are based on idealised models and are not necessarily representative of the structures found in real galaxies. It is clear, however, that differences in the stellar structure of infalling satellites may be an important factor in determining their stripping efficiency for at least part of an infalling satellite's lifetime and that the accuracy of the recovered stripping rates may also be influenced by these morphological differences.

Previous work by \citet{Chang2013} also finds that disc-dominated satellites are stripped more efficiently compared with spheroidal satellites once the tidal radius reaches the disc scale length. Although the regimes explored in this study differ significantly from \citet{Chang2013}, who study lower mass ratio mergers in less massive haloes, we still observe a similar relationship between satellite morphology and stripping efficiency. As seen in Figure \ref{fig:DM_stripping}, when the DM halo is poorly resolved, the DM half-mass radius quickly reaches the stellar half-mass radius, leading to a significant difference in the stripping efficiencies of discs and spheroids.

Regardless of morphology, mass loss is considerably more extreme for the most radial orbits and for the lowest masses, with satellites of $M_{\star} \leq 10^{9}$~M$_{\odot}$ on orbits with $r_{\rm peri}:r_{\rm apo}$ smaller than 1:10 all losing more than 50 per cent of their stars after 10 orbital periods in the benchmark run. For the lowest mass satellite and most radial orbit considered, satellites have been essentially completely stripped of stars after 10 orbital periods in the benchmark run.%At lower resolutions, stellar mass can be stripped even more quickly. In the most extreme cases essentially all stellar mass is stripped within five orbits for the most radial orbital configurations.

\begin{figure*}
    \centering
    \includegraphics[align=t, width=1\textwidth]{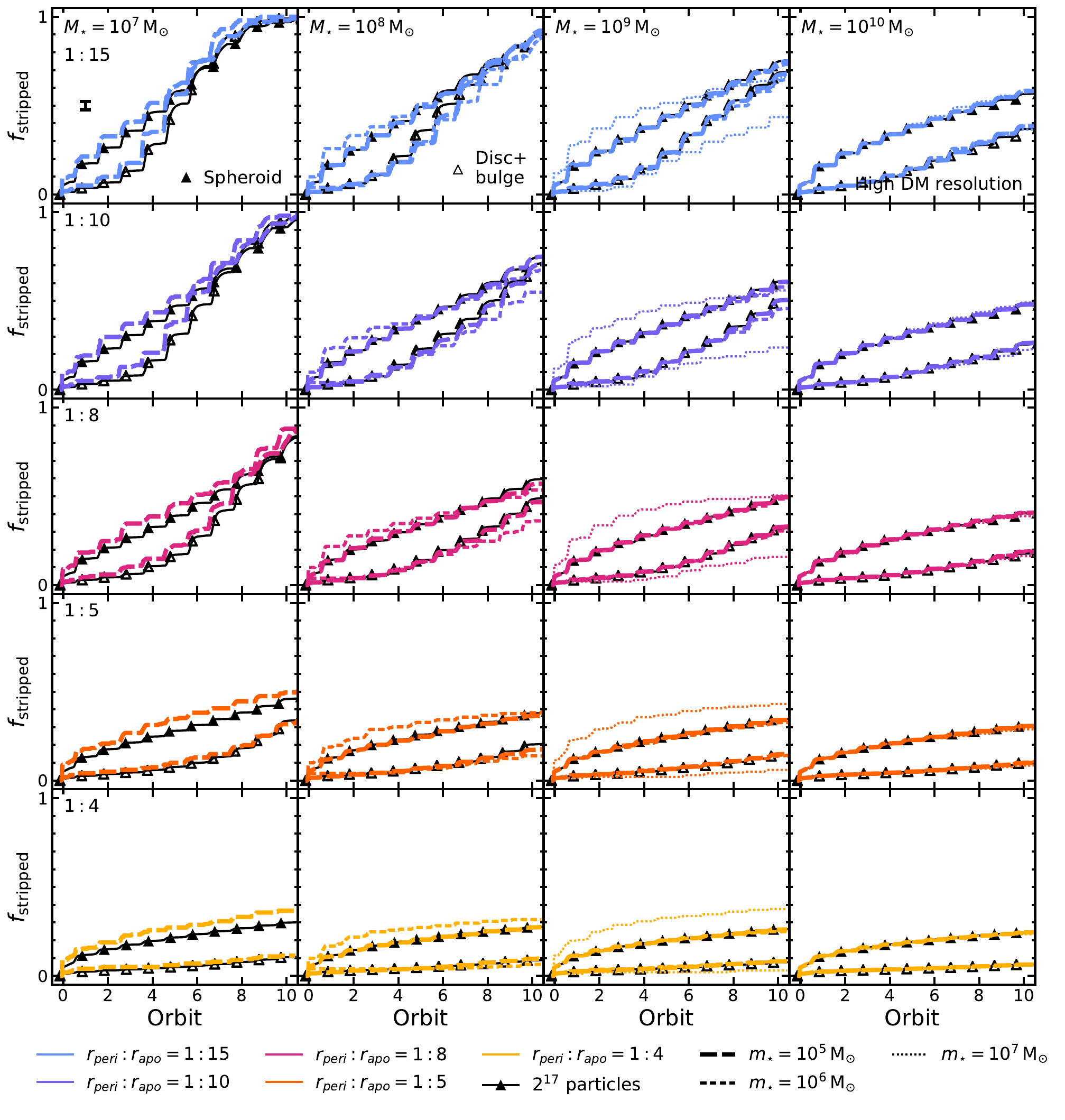}

    \caption{Tracks indicating the mass of stars stripped as a fraction of the initial mass for different mass satellites for the high DM resolution runs. More finely dashed, thinner lines indicate lower mass resolutions and the benchmark resolution is indicated by black lines with up-pointing triangle markers. Different colour lines indicate different orbital configurations. For the lowest resolution run, tracks shown as an average of 10 realisations, with the black error bar in the top-left panel indicating the variation in stripped fractions after 10 orbits over the 10 realisations averaged over the lowest resolution satellites at all masses.}
    \label{fig:stripping_tracks_sph_I}
\end{figure*}

Figure \ref{fig:stripping_tracks_sph_I} shows the mass of stars stripped as a fraction of the satellite's initial mass of each satellite as in Figure \ref{fig:stripping_tracks_sph_I_lr}, but for our high DM resolution runs. In this case, the resolution of the satellite DM halo is kept fixed at $2^{17}$ particles, ensuring it remains well resolved. We omit panels showing the highest mass satellites, as results are already converged for satellite stellar masses of $10^{10}$~M$_{\odot}$.

As discussed in Section \ref{sec:softening_stripping}, the significant over-stripping seen in Figure \ref{fig:stripping_tracks_sph_I_lr} is primarily driven by poorly resolved satellite inner halos. The much better agreement between the benchmark tracks at all but the lowest resolution levels in the high DM resolution runs seen in Figure \ref{fig:stripping_tracks_sph_I} indicates that very few ($N_{\star}\gtrsim 100$) star particles are necessary to capture stellar stripping rates with reasonable accuracy.

Even when the halo is appropriately resolved as is the case in the high DM resolution runs, we still see some deviation from the benchmark tracks at the lowest resolutions. Discretization noise plays a role in this. Poorly resolved satellites are expected to show more variance in stripping efficiency between orbits because the fraction of particles that are at large enough radii to become unbound during a given pericentric passage will fluctuate more significantly for objects resolved with fewer particles. This leads to both greater relaxation of the object (and therefore a more extended outer envelope of stars) as well as decreasing the average binding energies of particles, which can lead to a runaway amplification of the stripping rate which becomes more pronounced in stronger tidal fields \citep{vandenBosch2018}.

We investigate the effect of mass discretization by running 10 realisations of our simulations at the lowest resolution level. As indicated by the black error bars in the top-left panel of Figures \ref{fig:stripping_tracks_sph_I_lr} and \ref{fig:stripping_tracks_sph_I}, the typical variation in the stellar stripping efficiency across different realisations is small. Various studies have found that disruption may be a significant issue for DM subhaloes resolved with fewer than a few thousand particles \citep[][]{vandenBosch2018, Errani2021}. In our high DM resolution runs, we find that below a threshold of around 100 particles, divergence from the benchmark simulations becomes significant, but is still relatively modest compared with the divergence seen in the low DM resolution runs. The number of particles stripped during a given pericentric passage is typically very small, constituting a few tens of particles at the lowest resolution compared with thousands at the highest resolution. This can lead to increasingly over- or under-efficient stripping, which is amplified over time.

Once satellites are resolved with sufficiently few star particles we begin to see significant disruption in the stars, even though the DM halo is well resolved and remains undisrupted. While we do not observe comparable levels of runaway disruption in the satellite's stellar component in any of our set of low DM resolution run compared with those observed by \citet{vandenBosch2018} in the DM component, we are eventually able to produce a similar effect by reducing the stellar mass resolution even further.

We find that runaway stripping is easier to produce in higher mass satellites. For satellite stellar masses of $10^{11}$~M$_{\odot}$, rapid stripping proceeds once the massive satellite is resolved with a few hundred particles, whereas lower mass satellites require a significantly smaller number of particles before the same effect is observed. The higher levels of disruption seen in these satellites likely stems from their more extended profiles and higher stellar-to-halo-mass ratios, meaning particles are on average less bound and extend further from the centre of the cluster potential, therefore experiencing a stronger impulse at pericentre.

\subsection{Recovering bulk ICL contribution from stellar stripping}
\label{sec:recovering_ICL_fraction}

In this section, we explore how the total quantity of ICL originating from stripped stars depends on the DM and stellar resolution for resolutions typically achieved by contemporary cosmological simulations. 

\subsubsection{Trend with satellite stellar mass}

\begin{figure}
    \centering
    \includegraphics[width=0.45\textwidth]{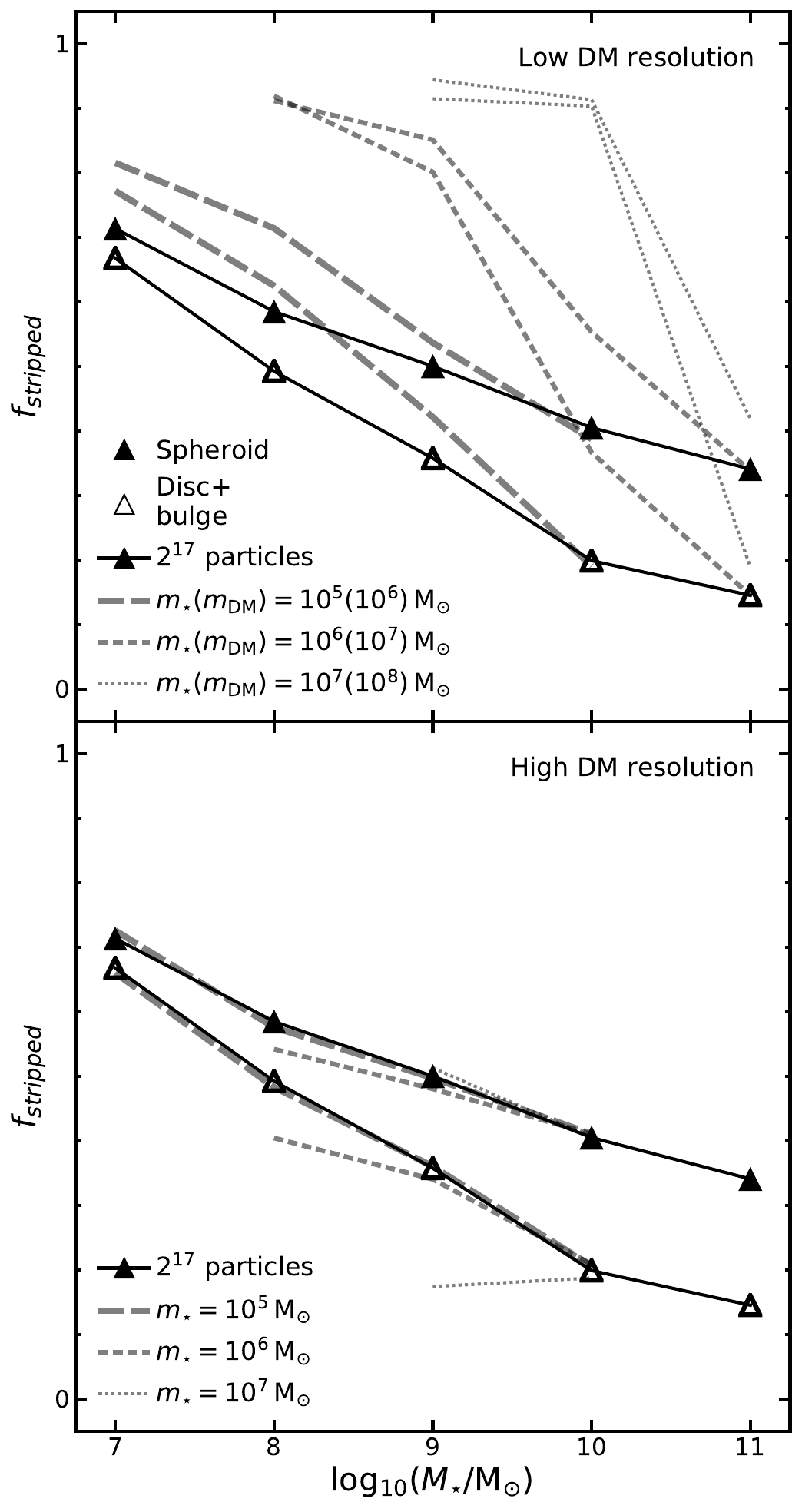}
    \caption{Comparison of total stellar mass stripped from satellites following 10 orbits as a function of satellite stellar mass for the low DM and high DM resolution runs presented as an average of all orbital configurations weighted according to their expected frequency from on \citet{Wetzel2011}. More finely dashed, thinner lines indicate lower mass resolutions and the benchmark resolution is indicated by black lines with up-pointing triangle markers. The top and bottom panels indicate the result for the low and high DM resolution runs respectively.}
    \label{fig:av_stripping_compare}
\end{figure}

We begin by summarising the stripped fractions of different mass satellites. Figure \ref{fig:av_stripping_compare} shows the total stellar mass stripped after 10 orbits as a function of the initial stellar mass of the satellite, with the top panel showing the result for the low DM resolution run and the bottom panel showing the result for the high DM resolution run. The results are presented as an average weighted by the expected frequency of different orbital configurations drawn from \citet{Wetzel2011}.

At the lowest resolution level, stripping rates in the low DM resolution run are significantly overestimated compared to the benchmark run, even at high masses. At higher resolutions, over-stripping remains significant for lower mass satellites, but approaches convergence at intermediate masses ($M_{\star} \geq 10^{10}$~M$_{\odot}$).

In the high DM resolution run, results are well converged compared with the benchmark for all but the lowest resolution level. For all resolution levels, there is some under- or over-stripping when satellites are resolved with very few particles. This appears to become more pronounced in higher mass satellites than in lower mass ones.

\subsubsection{Total stellar mass stripped}

We now combine the stripped fractions shown in Figure \ref{fig:av_stripping_compare} with a galaxy stellar mass function (GSMF) described by a \citet{Schechter1976} function with a characteristic mass ($M^{\star}$) of $10^{11}$~M$_{\odot}$ and low-mass slope ($\alpha$) of -1.4, typical for the field galaxy population \citep[e.g.][]{Sedgwick2019}. Results beyond $10^{11}$~M$_{\odot}$ and below the resolution limit (of 100 star particles) for a given mass resolution are extrapolated as a power-law up to $10^{12}$~M$_{\odot}$ and down to a stellar mass equivalent to 20 particles. Note that the extrapolation down to lower masses does not significantly impact the quantity of ICL predicted. If we do not perform this extrapolation, the total mass of ICL predicted varies by a few per cent in the worst case. 

\begin{figure}
    \centering
    \includegraphics[width=0.45\textwidth]{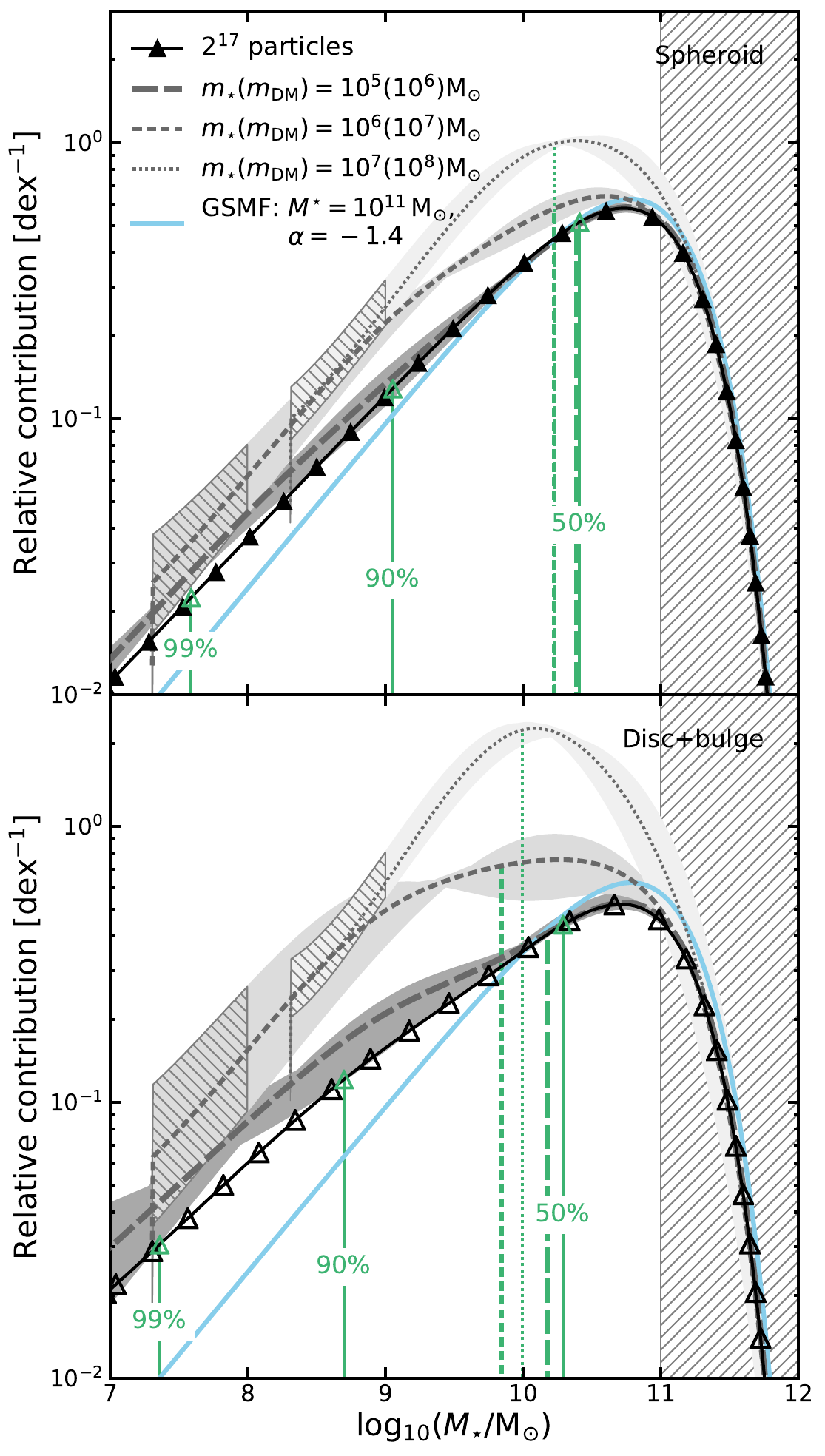}
    \caption{The contribution of satellites of different stellar masses to the bulk mass of the ICL, normalised relative to the benchmark simulation. Grey lines show the average contribution to the ICL mass budget due to stellar stripping from satellites of a given mass, weighted by the expected frequency of different orbital configurations from \citet{Wetzel2011}. More finely dashed lines indicate poorer resolution runs and black lines with up-pointing triangles indicate the benchmark run. Filled regions indicate the range of values over all orbital configurations for each resolution, with lighter filled regions corresponding to poorer resolution. Results beyond $10^{11}$~M$_{\odot}$ and below the resolution limit (of 100 star particles) for a given mass resolution are extrapolated as a power-law up to $10^{12}$~M$_{\odot}$ and down to the mass equivalent to 20 particles as indicated by hatched regions. The pale blue line shows the relative contribution of satellites of different masses to the overall mass budget of the GSMF (i.e. the relative contribution of the stripping efficiency is flat with satellite mass). Solid green lines with arrows mark satellite masses above which 50, 90 and 99 per cent of ICL mass is accounted for in the benchmark run, with different style dashed green lines indicating the same for their corresponding resolution run at the 50 per cent level.}
    \label{fig:budget_contribution}
\end{figure}

Figure \ref{fig:budget_contribution} shows the expected contribution to the ICL from satellites of different masses with spheroidal morphologies (top panel) and disc-dominated morphologies (bottom panel). Lower resolution runs are indicated by increasingly finely dashed lines and the benchmark run is indicated by solid black lines with up-pointing triangular markers. Lines are normalised relative to the benchmark run, so that a larger area under the curve corresponds to a greater quantity of mass stripped in total. Filled regions indicate the range of values over all orbital configurations and the hatched portions of each filled region indicate where we have extrapolated our results from 100 down to 20 particles. The larger hatched region indicates masses above which we extrapolate our results\footnote{We note that, in this mass regime, mergers between satellites and the BCG are likely to begin playing a significant role in the production of ICL. These processes are ignored in this work.}. 

Labelled green arrows with solid lines mark masses above which 50, 90 and 99 per cent of ICL mass is accounted for in the benchmark run and dashed green lines indicate the 50 per cent value for each corresponding resolution level. We also indicate the relative contribution to the total stellar mass budget for the adopted GSMF using pale blue lines. 

In common with a number of observational \citep{DeMaio2018,Montes2018,Montes2021} and theoretical \citep{Contini2014,Chun2023,Ahvazi2024,Brown2024} results, we find that the peak contribution to the ICL from stripped stars comes from intermediate mass satellites (between $10^{10}$~M$_{\odot}$ and $10^{11}$~M$_{\odot}$). This is true of all resolution levels, although the lowest resolution runs predict a significantly greater relative contribution from galaxies in this range compared with lower mass satellite galaxies.

We observe relatively small differences between the shape (but significant differences in normalisation) of the relative ICL contribution and the shape of the relative total contribution to the GSMF (shown in blue). Because lower mass satellites are more efficiently stripped, we observe an excess in the relative stripped mass contribution compared with the GSMF at the low-mass end compared with a small deficiency at the peak. However, the overall shape of the distribution is clearly determined primarily by the shape of the GSMF.

Because the GSMF varies by orders of magnitude across the mass range, while the stripped fraction can vary by unity at most, it is difficult to considerably change the shape of Figure \ref{fig:budget_contribution} unless stripping efficiency were suppressed significantly at higher masses so that it was very close to zero. In reality, it may not be possible to reproduce observed ICL fractions with low-mass galaxies alone. For example, galaxies less massive than $10^{9.5}$~M$_{\odot}$ constitute around 14 per cent of the total mass budget under the GSMF, so even if their stripping efficiency is close to unity \citep[incompatible with observations given that the observed difference between the cluster and field galaxy stellar mass functions are too modest;][]{Vulcani2013,vanderBurg2018}, there is not sufficient mass contained within low-mass galaxies to account for all of the ICL, even under the conservative assumption that ICL accounts for 10 to 20 per cent of the total stellar mass of the cluster.

Is also worth noting that, since the ICL mass budget contribution follows the GSMF quite closely, the peak contribution (between $10^{10}$~M$_{\odot}$ and $10^{11}$~M$_{\odot}$) also corresponds roughly to the knee of stellar-to-halo mass relation \citep{Moster2013}. As a result, the objects that originate the bulk of the ICL are typically resolved with the fewest DM particles relative to their stellar mass. As previously discussed in Section \ref{sec:fixed_n}, stellar mass resolution is of minor importance compared the DM resolution, meaning considerably higher resolution is required to accurately recover stripping rates than would be necessary if ICL originated primarily from low-mass galaxies. Because less DM resolution is needed to properly resolve galaxies either side of the knee of the stellar-to-halo mass relation, more efficient alternatives to fixed mass resolution in cosmological simulations could include dynamically splitting and merging particles based on halo mass or replacing infalling haloes with sufficiently high-resolution models, similar to the approaches employed by \citet{Vacondio2013} or \citet{Chun2022}.

\begin{figure}
    \centering
    \includegraphics[width=0.45\textwidth]{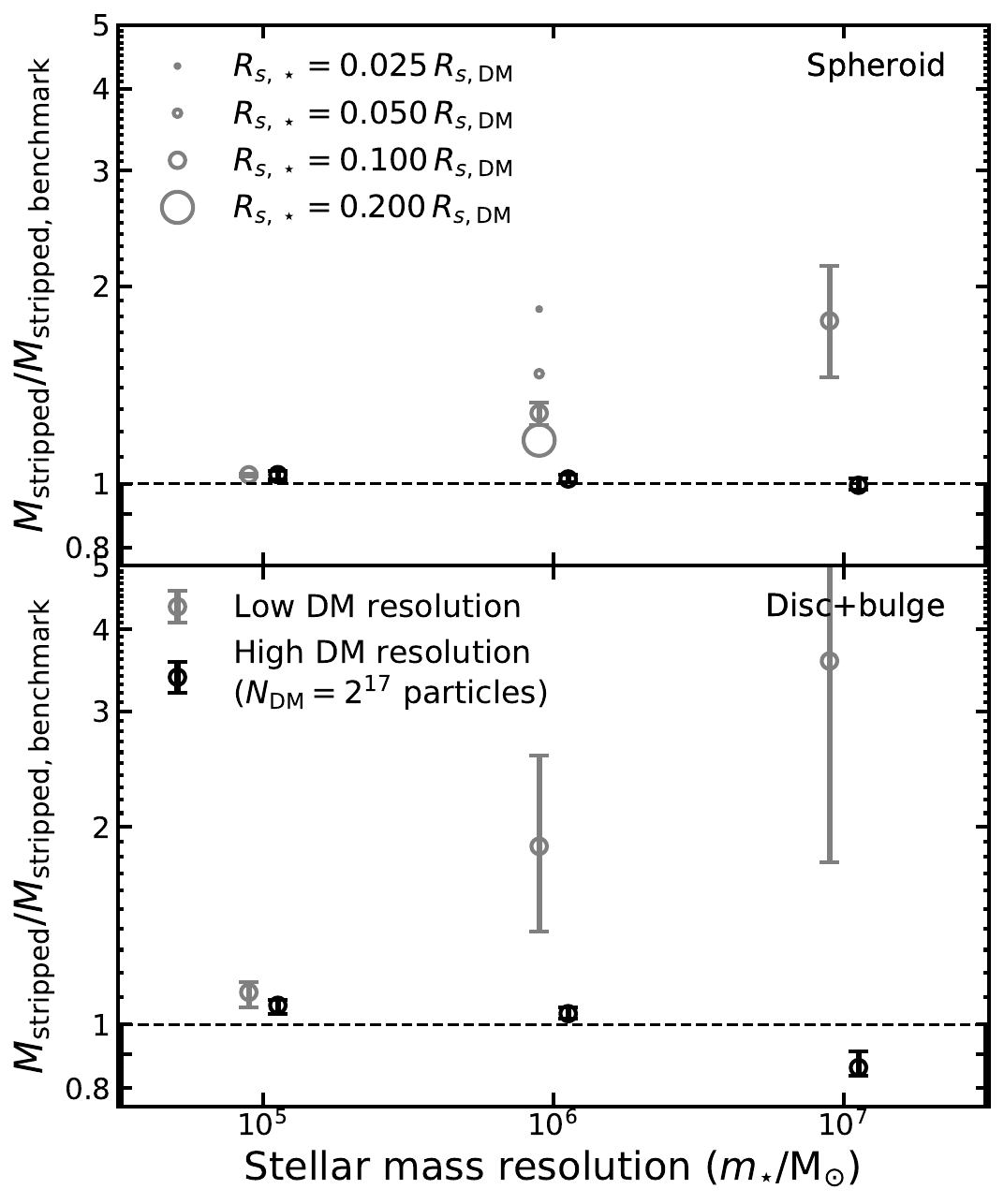}
    \caption{The total mass of stars stripped following 10 orbits in the low and high DM resolution runs relative to the benchmark run. Low and high DM runs are indicated by grey and black markers respectively (see Table \ref{tab:resolution} for details). Error bars indicate the range of values obtained across all orbital configuration. Additional grey markers are shown for the low DM resolution run with $m_{\star}=10^{6}$~M$_{\odot}$ stellar mass resolution, with larger markers indicating larger satellite scale lengths relative to the halo scale length.}
    \label{fig:cumulative_budget}
\end{figure}

Figure \ref{fig:cumulative_budget} shows the total stellar mass stripped over 10 orbits relative to the benchmark run following the same assumptions (mass function and distribution of orbital configurations) as above. Grey points show results for the same low DM resolution runs as shown in Figure \ref{fig:budget_contribution} and black points show the same for the high DM resolution runs. Error bars indicate the range of values obtained individually across all orbital configurations.

In the low DM resolution runs, we see that the total stripped stellar mass predicted increases significantly towards poorer resolution, especially in the case of disc-dominated galaxies. Whereas, for the high DM resolution runs, results are consistently well converged compared with the benchmark run. 

In reality, the mass distribution of galaxies that contribute to ICL production is not described by a single GSMF as the ICL is formed by galaxies that fall into the cluster at different redshifts. We explore how sensitive our results are to the shape of the GSMF by considering a range of low-mass slopes between $\alpha=-1.6$ and $\alpha=-0.5$. In all cases, more positive values of $\alpha$ result in better convergence with the benchmark results, because a more positive slope implies fewer low-mass satellites, which are most poorly resolved. However, over a realistic range of $\alpha$, we do not observe any qualitative change in our results.

Finally, we consider the effect of galaxy sizes on the level of convergence with the benchmark run. In addition to expected variations in galaxy size as a function of redshift or morphology, different approaches to calibration may also produce differences in galaxy scaling relations across simulations, including the size-mass relation \citep[e.g.][]{Crain2015,Dubois2016,Pillepich2018b}. Even where simulations are calibrated to reproduce $z=0$ scaling relations, they may not be consistent with observed low-mass galaxies (Watkins et al., in prep; Martin et al., in prep) or match observed evolutionary trends with redshift \citep[e.g.][]{Parsotan2021}.

For the low DM resolution run with a stellar and DM mass resolution of $10^{6}$~M$_{\odot}$, $10^{7}$~M$_{\odot}$ respectively, we show results for satellites with stellar components double, half and one quarter the size of our standard spheroid model, indicated as different size circles. As expected, based on the results of Section \ref{sec:softening_stripping}, we see that our results are not as well converged for smaller satellites, since the stellar mass of smaller galaxies is distributed closer to the centre of the potential where the DM density profile becomes unresolved.

\subsubsection{Comparison with cosmological simulations}

Our idealised simulations imply that considerably higher resolutions may be necessary to properly resolve the ICL than some previous studies. For example, \citet{Puchwein2010} demonstrated that stellar mass, DM mass and spatial resolutions of around $10^{7}$~M$_{\odot}$, $10^{8}$~M$_{\odot}$ and 3~kpc respectively resulted in convergence of the bulk quantity of ICL produced in their cosmological zoom-in hydrodynamical simulations of clusters with AGN and stellar feedback. Similarly, \citet{Pillepich2018b} also claim no significant difference in bulk ICL \textit{fractions} recovered by the Illustris TNG100 and lower-resolution TNG300 simulations. However, they note that the galaxy evolution model of the TNG suite of simulations is calibrated to reproduce observed relations only at intermediate TNG100 resolution, so galaxy statistics such as the galaxy mass function and stellar-to-halo mass relation are not fully converged in the lower resolution simulation.

We also compare our results with clusters from TheThreeHundred Project \citep{Cui2018} by measuring ICL masses from TheThreeHundred GIZMO runs \citep{Hopkins2015, Dave2019, Cui2022}, which simulate 324 clusters at a (maximum) stellar mass and DM mass resolution of $m_{\star}=2.4\times10^{8}$~$h^{-1}$~M$_{\odot}$ and $m_{\rm DM}=1.5\times10^{9}$~$h^{-1}$~M$_{\odot}$ \citep[GIZMO 3k;][]{Cui2022} and $m_{\star}=3\times10^{7}$~$h^{-1}$~M$_{\odot}$ and $m_{\rm DM}=1.8\times10^{8}$~$h^{-1}$~M$_{\odot}$ (GIZMO 7k; Cui et al. in prep).
%$m_{\star}=2.4\times10^{8}$~$h^{-1}$M$_{\odot}$ and $m_{DM}=1.5\times10^{9}$~$h^{-1}$M$_{\odot}$ (GIZMO 7k). 

\begin{figure}
    \centering
    \includegraphics[width=0.45\textwidth]{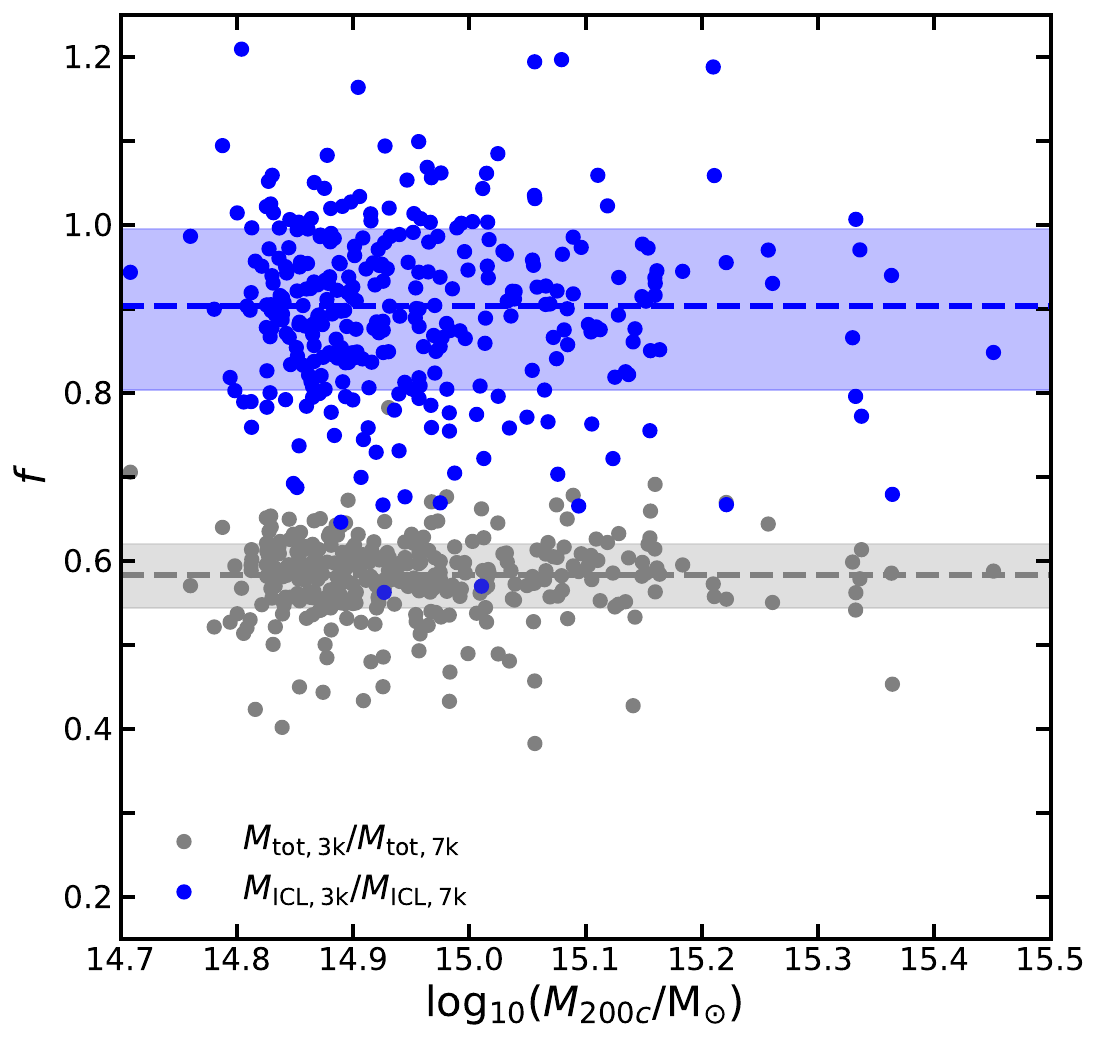}
    \caption{Ratio of quantities measured between TheThreeHundred Project GIZMO 3k and GIZMO 7k runs. Grey points indicate the ratio of total stellar masses within $R_{200c}$ and blue points indicate the ratio of ICL masses within $R_{200c}$. Dashed lines and filled regions indicate median values and $1\sigma$ dispersion respectively.}
    \label{fig:ThreeHundred}
\end{figure}

For each cluster, we measure the total stellar mass and total ICL mass within $R_{200c}$ following \citet{Contreras2024}. We use a 70~kpc aperture to exclude the BCG, but note that using an aperture of 30, 50, 70 and 200~kpc does not qualitatively alter our results.

In Figure \ref{fig:ThreeHundred}, we show how the total stellar mass and total ICL mass differs between the two different resolution runs. Grey points show the ratio of total stellar masses measured between the same clusters in the GIZMO 3k and GIZMO 7k runs, while blue points show the ratio for ICL masses. Dashed lines and filled regions indicate the median and $1\sigma$ dispersion for each quantity. ICL masses are only slightly smaller on average in the 3k runs, despite cluster total stellar masses being around 60 per cent that of their high-resolution counterparts. This indicates that stellar stripping must be roughly two thirds more efficient than in the GIZMO 3k run in order to account for the smaller reservoir of stellar mass available to be stripped.

Our results based on TheThreeHundred simulation are in qualitative agreement with those from our idealised simulations, but there appears to be some tension with those of \citet{Puchwein2010} and \citet{Pillepich2018b}. Given that our simulations are idealised and that we only consider stellar stripping, it is important to note that there are several caveats of this study which may cause discrepancies to arise:

\begin{enumerate}
    \item Firstly, although stripping is expected to be the dominant channel for ICL formation \citep{Contini2014}, our approach neglects potentially more resolution-agnostic formation channels modelled ab initio, such as by \citet{Puchwein2010}, \citet{Pillepich2018b} and other studies which utilise cosmological simulations. For example, stars may enter the ICL via violent relaxation \citep{LyndenBell1967,Albada1982} during galaxy mergers rather than as a response to a more gradually varying tidal field or may be deposited via pre-processing rather than being stripped directly within the cluster. For the satellite masses considered here, pre-processed stars likely represent only a few per cent of total satellite stellar masses at infall \citep{Martin2022,Proctor2024} but mergers of clusters and groups are expected to bring in much more significant quantities of pre-processed material \citep{Mihos2017,Contini2024}.
    \item Additionally, some theoretical studies find up to 30 per cent of ICL stars forming directly in-situ, (i.e. stars forming at very large radii from galaxies rather than within galaxies). Although the compatibility of this channel with observations is disputed \citep[e.g.][]{Melnick2012,Contini2024}, it has been shown to be present in some cosmological simulations across a wide range of resolutions including by \citet{Puchwein2010} and in Illustris TNG simulations \citep[e.g.][]{Ahvazi2024}. In these studies, in-situ formation represents a significant and possibly resolution-independent channel for ICL production. We note that, in contrast, in-situ ICL formation accounts for an average of only 1 per cent in TheThreeHundred GIZMO 3k run (Contreras-Santos, in prep). This may at least partially explain why \citet{Pillepich2018b} report that improving mass resolution results in no change to their ICL fractions, while our results using TheThreeHundred simulation indicate decreasing ICL fractions.
    \item Our study only includes satellite haloes with a \citet{Hernquist1990} density profile, but wide diversity in halo mass profiles has been measured observationally \citep[e.g.][]{Oh2015}. This diversity has been shown to emerge ab initio in high-resolution simulations as a result of genuine astrophysical mechanisms \citep[e.g. cored haloes as a result of feedback; ][]{Jackson2024}. An inability to resolve the inner radii of galaxy haloes at lower resolutions may therefore have a less significant impact on numerical convergence if a significant fraction of galaxies would have hosted genuine flat inner density profiles at higher resolution.
    \item Similarly, significant diversity is observed in galaxy morphologies and sizes beyond the two simple models used in this study \citep[e.g.][]{Kelvin2014,Lazar2024}. Although we have briefly investigated the effect of galaxy size on convergence, we have not considered how varying the morphological mix or galaxy sizes with mass may effect this.
\end{enumerate}

\subsection{Which galactic radii are stars stripped from?}
\label{sec:location}

Finally, we consider how the satellite galactic radii from which stellar mass is stripped depend on resolution.

\begin{figure*}
    \centering
    \includegraphics[width=0.95\textwidth]{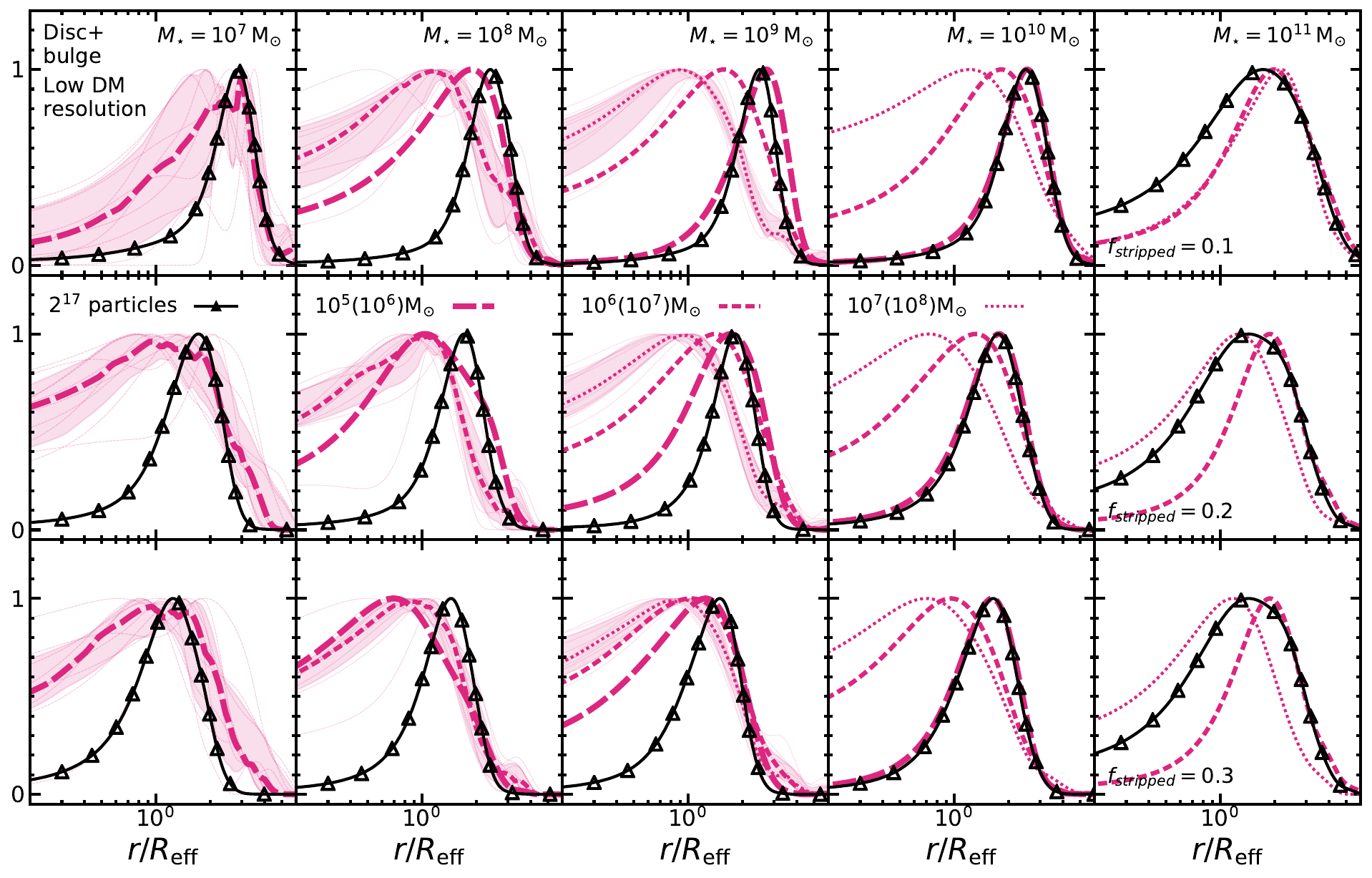}
    \caption{The radial distribution of stars stripped from satellites with disc-dominated morphologies at a given value of $f_{\rm stripped}$ for the same 1:8 $r_{\rm peri}:r_{\rm apo}$ orbital configuration for the low DM resolution run. More finely dashed and thinner lines indicate lower resolutions and black likes with up-pointing triangle markers indicate the benchmark run. The legend, split across the middle panels, indicates the corresponding stellar mass resolution with the DM mass resolution in parentheses. For the least resolved satellites, consisting of 100 star particles, filled regions indicate the range of values over 10 realisations, with thin lines indicating radial distributions for individual realisations.}
    \label{fig:stripping_radius_detail_lr}
\end{figure*}

Figure \ref{fig:stripping_radius_detail_lr} shows the distribution of radii from which stars are stripped from disc-dominated satellites in the low DM resolution runs. The results we obtain for spheroidal morphologies are qualitatively similar. We calculate the radial distribution of star particles based on their radius at $t=0$ rather than at the time of stripping. Since resolution also influences the stripping rate, we present distributions at different times corresponding to a given value of $f_{\rm stripped}$ rather than at a specific time.

We show the distribution of stripping locations for a single orbital configuration ($r_{\rm peri}:r_{\rm apo}=1:8$) for satellites with disc-dominated morphologies. More finely dashed lines indicate poorer resolution runs and filled regions indicate the range of values obtained from 10 realisations of the lowest resolution run in each mass bin. Black lines with up-pointing triangle markers indicate the benchmark run. From top-to-bottom, panels indicate distributions for values of $f_{\rm stripped}$ of 0.1, 0.2 and 0.3, as indicated in the rightmost panel of each row. From left-to-right panels indicate the radial distributions for different mass satellites indicated in the topmost panel of each column.

In lower resolution runs, stars originating from smaller radii at $t=0$ end up being stripped considerably sooner than they would be in the benchmark runs. Even for satellites whose stellar stripping tracks that appear relatively well resolved in Figure \ref{fig:stripping_tracks_sph_I_lr}, we still observe quite significant differences in terms of where these stripped stars originate compared with the benchmark run.

For brevity, we do not show results for the high DM resolution runs here but note that we observe a similar trend for stars originating from smaller radii being stripped earlier towards lower resolution. However, the level of deviation from the benchmark runs is relatively slight and considerably less extreme compared with the low DM resolution runs. This indicates that the increase in overstripping from the central regions of the satellite towards lower resolutions seen in Figure \ref{fig:stripping_radius_detail_lr} is caused by a combination of the migration of particles to larger radii due to the poorly resolved inner DM halo profile (discussed in Section \ref{sec:fixed_n}) and discretization noise.

\begin{figure*}
    \centering
    \includegraphics[width=1\textwidth]{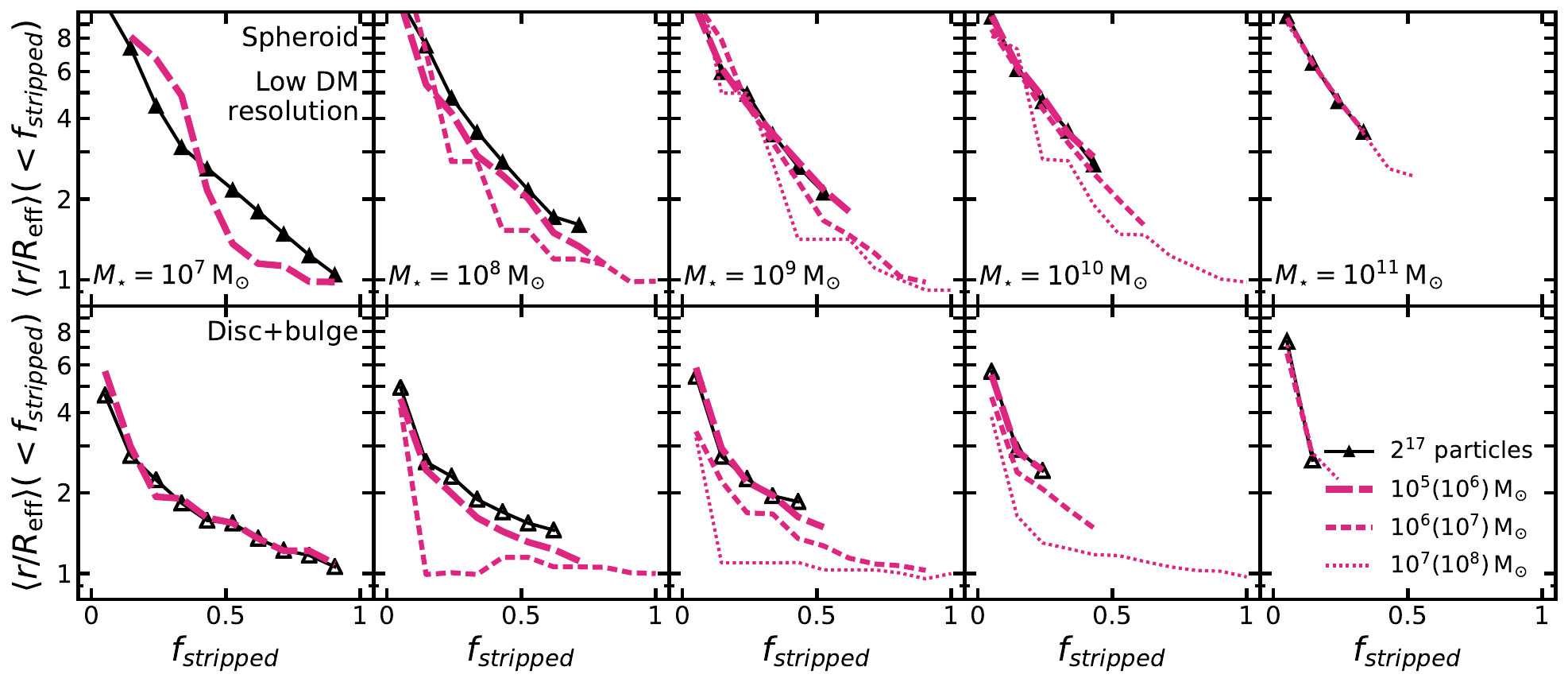}
    \caption{Tracks showing the average radius from which stars are stripped as a function of stripped fraction for the same 1:8 $r_{\rm peri}:r_{\rm apo}$ orbital configuration as in Figure \ref{fig:stripping_radius_detail_lr} for the low DM resolution runs. The top and bottom set of panels show our results for spheroidal and disc-dominated satellites respectively. More finely dashed and thinner lines indicate lower resolutions and the benchmark resolution is indicated by black lines with up-pointing triangle markers. The legend in the bottom-right panel indicates the corresponding stellar mass resolution with the DM mass resolution in parentheses. For the least resolved satellites, consisting of 100 star particles, tracks are an average of 10 realisations.}
    \label{fig:stripping_radius_I_lr}
\end{figure*}

Finally, we summarise the evolution of the stellar stripping radius, which we define here as the median radius from which stars are stripped at a given value of $f_{\rm stripped}$ (again using their radius at $t=0$ in order to control for different galaxy size evolution in different resolution runs). Figure \ref{fig:stripping_radius_I_lr} shows the stripping radius averaged over all stripped particles prior to reaching a given value of $f_{\rm stripped}$ in the low DM resolution runs. Different resolutions are indicated by increasingly finely dashed lines for lower resolution and the benchmark run is indicated by black lines with up-pointing triangle markers. 

We again see that stars are stripped from much closer to the centre for the lowest resolution runs. For clarity, we do not plot our results for all orbital configurations. However, we note that we do not observe a strong dependence on the orbital configuration. 

Because the chemodynamical properties of stars that make up satellites vary with radius from their centre, the order in which different components are stripped is important for determining the resolved properties of the ICL, for example its observed colour, age or metallicity gradients \citep[e.g.][]{DeMaio2018,Montes2018,Montes2021,Marx2023}. Without sufficient DM and stellar resolution, stars are stripped both at faster rates and from the wrong locations within the satellite, which may lead to significant differences not only in the bulk quantity of ICL but also in its resolved properties.

\section{Summary}
\label{sec:summary}

In this study, we presented a suite of idealised N-body simulations designed to explore the effect of numerical resolution on the stripping efficiency of satellites in a static cluster potential. We probed stripping efficiency as a function of satellite mass, morphology and orbital configuration for simulations run with a range of numerical resolutions and force softening lengths. Our main conclusions are as follows:

\begin{enumerate}
\item \noindent \textit{The properties of the satellite DM halo are one of the primary drivers of the stellar stripping efficiency.} In particular, the slope of the inner DM profile close to the extent of the stellar component strongly influences the rate at which stars are stripped.
\\
\item \noindent \textit{Poorly resolved satellite DM haloes develop cored inner profiles, making stellar stripping much more efficient.} Although the overall mass and size evolution of satellite haloes remain relatively robust at each resolution level, we observe a significant divergence in their inner profiles, which can evolve to become almost completely flat out past the extent of the stellar component, even when the DM halo is resolved with more than 1000 particles. This leads to large spurious increases in the stellar stripping rate due to both the shallower central potential and spurious expansion of the stellar component. This effect can be the result of either insufficient numerical resolution in the DM or using force softening lengths that are comparable to or larger than the extent of the stellar component of the satellite galaxy. The halo must be well sampled and well resolved spatially down to the radius of the stellar component to recover accurate stripping rates.
\\
\item  \noindent \textit{The stellar stripping tracks of satellites with well resolved DM haloes are converged once their stellar component is resolved with between a few hundred and one thousand star particles.} Discretization noise is responsible for a small level of variance in the tracks which is amplified over time. A more extreme consequence of discretization occurs when this amplification leads to runaway disruption of the stellar component. However, this only becomes an issue for the most massive satellites in our simulations $(M_{\star} = 10^{11}$~M$_{\odot})$, which experience a stronger tidal field at pericentric passage due to their larger extent. However, given their large stellar masses, almost all contemporary cosmological simulations have sufficient resolution to prevent this in practice.
\\
\item  \noindent \textit{When adopting a range of DM, stellar mass and spatial resolutions typical of contemporary cosmological simulations lower masses and lower resolution satellites are increasingly over-stripped compared to the benchmark resolution.} Although they typically remain converged for at least a few orbits, stellar stripping tracks eventually diverge from the benchmark. A stellar mass and DM mass resolution of $m_{\star}=10^{5}$~M$_{\odot}$ and $m_{\rm DM}=10^{6}$~M$_{\odot}$, is required for tracks to remain relatively well converged for 10 orbits ($\sim 5$~Gyrs) across the full range of satellite masses explored ($10^{7} \leq $$M_{\star}$/M$_{\odot} \leq 10^{11}$), although we still observe some modest over-stripping at lower masses ($M_{\star} < 10^{9}$M$_{\odot}$).
\\
\item \noindent \textit{Typical resolutions achieved by contemporary cosmological simulations may significantly overestimate the bulk quantity of stellar mass liberated due to stellar stripping.} Assuming a galaxy stellar mass function with a characteristic mass of $M^{\star}=10^{11}$~M$_{\odot}$ and low-mass slope of $\alpha=-1.4$, and assuming that all satellites survive to complete 10 orbits without a significant change in their pericentric radius, we find that simulations with stellar mass and DM mass resolutions of $m_{\star}=10^{5},10^{6},10^{7}$~M$_{\odot}$ and $m_{\rm DM}=10^{6},10^{7},10^{8}$~M$_{\odot}$ are expected to produce a 3, 28 and 66 per cent excess in the amount of stellar mass stripped compared to the benchmark for satellites with spheroidal morphologies. We observe a significantly larger excess for disc-dominated galaxies. However, in practice, at masses where stripping tracks begin to diverge, numerical resolutions are likely too low to allow cosmological simulations to form discs.
\\
\item  \noindent \textit{A majority of stellar mass liberated due to stellar stripping only is expected to originate in satellites with intermediate stellar masses.} Assuming a galaxy stellar mass function with the same characteristic mass and low-mass slope as above and assuming all satellites survive on the same orbits, we expect a majority of the mass liberated to move to the ICL as a result of stellar stripping to originate in galaxies with infall masses between $10^{10}$~M$_{\odot}$ and $10^{11}$~M$_{\odot}$. Lower resolution runs produce peak contributions at progressively lower stellar masses compared with the benchmark run, peaking at stellar masses of $5\times 10^{10}$, $4\times 10^{10}$ and  $2\times 10^{10}$ for stellar mass and DM mass resolutions of $m_{\star}=10^{5},10^{6},10^{7}$~M$_{\odot}$ and $m_{\rm DM}=10^{6},10^{7},10^{8}$~M$_{\odot}$ respectively.
\\
\item  \noindent \textit{The radius within the satellite from which stars are stripped is sensitive to both stellar and DM resolution.} Stars are preferentially stripped from the centre of the satellite at low-resolutions. This can be due to both runaway disruption of the stellar component and due to the halo being too poorly resolved in the centre. This has implications for the recovery of the resolved properties of ICL, since the chemodynamical properties of stars are expected to vary across satellite galaxies. Although low-mass satellites contribute relatively little to the total stellar mass budget of the ICL, the fact that they are both over-stripped and experience stripping at spuriously small radii at low-resolutions may result in inaccurate resolved properties of the ICL, given that lower mass objects are expected to increase their contribution to the ICL mass budget towards the outskirts of the cluster \citep[e.g.][]{Rudick2010,Kluge2024}.
\end{enumerate}

Our results show that many state-of-the-art cosmological simulations in use today may overestimate stellar stripping rates, even in satellites of relatively high stellar mass. Since insufficient resolution also results in stars being stripped from the inner regions of satellites very early on, this may not just lead to differences in the bulk quantity of the ICL, but also its resolved chemodynamical properties. Although it is possible that these problems may be diminished, for example if more resolution agnostic processes such as violent relaxation are responsible for producing a significant quantity of the ICL, our findings demonstrate a need to exercise caution when interpreting predictions from cosmological simulations, particularly for objects with DM haloes resolved with fewer than $\sim 10,000$ particles.

%................................................................................................

\section*{Acknowledgements}

G.~M, F.~R.~P, and N.~A.~H acknowledge support from the UK STFC under grant ST/X000982/1. N.~A.~Hatch gratefully acknowledges support from the Leverhulme Trust through a Research Leadership Award. A.~K is supported by the Ministerio de Ciencia e Innovaci\'{o}n (MICINN) under research grant PID2021-122603NB-C21. W.~C is supported by the Atracci\'{o}n de Talento Contract no. 2020-T1/TIC-19882 was granted by the Comunidad de Madrid in Spain, and the science research grants were from the China Manned Space Project. W.~C also thanks the Ministerio de Ciencia e Innovación (Spain) for financial support under Project grant PID2021-122603NB-C21 and HORIZON EUROPE Marie Sklodowska-Curie Actions for supporting the LACEGAL-III project with grant number 101086388. 

The research in this paper made use of the SWIFT open-source simulation code \citep[http://www.swiftsim.com, ][]{Schaller2018} version 1.0.0.

Part of this work has been made possible by TheThreeHundred collaboration, which benefits from financial support of the European Union’s Horizon 2020 Research and Innovation programme under the Marie Skłodowskaw-Curie grant agreement number 734374, i.e. the LACEGAL project. TheThreeHundred simulations used in this paper have been performed in the MareNostrum Supercomputer at
the Barcelona Supercomputing Center, thanks to CPU time granted by the Red Española de Supercomputación. The high-resolution (7K) simulations from TheThreeHundred were performed on multiple Supercomputers: MareNostrum, Finisterrae3, and Cibeles through The Red Española de Supercomputación grants, DIaL3 -- DiRAC Data Intensive service at the University of Leicester through the RAC15 grant, and the Niagara supercomputer at the SciNet HPC Consortium.

We are grateful to Jesse Golden-Marx and Joseph Butler for helpful comments.

\section*{Data Availability}

Data will be shared on request to the corresponding author. Some of the results shown in this work use data from TheThreeHundred galaxy clusters sample. These data are available on request following the guidelines of TheThreeHundred collaboration, at https://www.the300-project.org. 

%%%%%%%%%%%%%%%%%%%%%%%%%%%%%%%%%%%%%%%%%%%%%%%%%%



\bibliographystyle{mnras}
\bibliography{paper_mnras} % if your bibtex file is called example.bib

\begin{thebibliography}{}
\makeatletter
\relax
\def\mn@urlcharsother{\let\do\@makeother \do\$\do\&\do\#\do\^\do\_\do\%\do\~}
\def\mn@doi{\begingroup\mn@urlcharsother \@ifnextchar [ {\mn@doi@} {\mn@doi@[]}}
\def\mn@doi@[#1]#2{\def\@tempa{#1}\ifx\@tempa\@empty \href {http://dx.doi.org/#2} {doi:#2}\else \href {http://dx.doi.org/#2} {#1}\fi \endgroup}
\def\mn@eprint#1#2{\mn@eprint@#1:#2::\@nil}
\def\mn@eprint@arXiv#1{\href {http://arxiv.org/abs/#1} {{\tt arXiv:#1}}}
\def\mn@eprint@dblp#1{\href {http://dblp.uni-trier.de/rec/bibtex/#1.xml} {dblp:#1}}
\def\mn@eprint@#1:#2:#3:#4\@nil{\def\@tempa {#1}\def\@tempb {#2}\def\@tempc {#3}\ifx \@tempc \@empty \let \@tempc \@tempb \let \@tempb \@tempa \fi \ifx \@tempb \@empty \def\@tempb {arXiv}\fi \@ifundefined {mn@eprint@\@tempb}{\@tempb:\@tempc}{\expandafter \expandafter \csname mn@eprint@\@tempb\endcsname \expandafter{\@tempc}}}

\bibitem[\protect\citeauthoryear{{Adami}, {Biviano}, {Durret}  \& {Mazure}}{{Adami} et~al.}{2005}]{Adami2005}
{Adami} C.,  {Biviano} A.,  {Durret} F.,   {Mazure} A.,  2005, \mn@doi [\aap] {10.1051/0004-6361:20053504}, \href {https://ui.adsabs.harvard.edu/abs/2005A&A...443...17A} {443, 17}

\bibitem[\protect\citeauthoryear{{Ahvazi}, {Sales}, {Doppel}, {Benson}, {D'Souza}  \& {Rodriguez-Gomez}}{{Ahvazi} et~al.}{2024}]{Ahvazi2024}
{Ahvazi} N.,  {Sales} L.~V.,  {Doppel} J.~E.,  {Benson} A.,  {D'Souza} R.,   {Rodriguez-Gomez} V.,  2024, \mn@doi [\mnras] {10.1093/mnras/stae848}, \href {https://ui.adsabs.harvard.edu/abs/2024MNRAS.529.4666A} {529, 4666}

\bibitem[\protect\citeauthoryear{{Arnaboldi}, {Gerhard}, {Aguerri}, {Freeman}, {Napolitano}, {Okamura}  \& {Yasuda}}{{Arnaboldi} et~al.}{2004}]{Arnaboldi2004}
{Arnaboldi} M.,  {Gerhard} O.,  {Aguerri} J. A.~L.,  {Freeman} K.~C.,  {Napolitano} N.~R.,  {Okamura} S.,   {Yasuda} N.,  2004, \mn@doi [\apjl] {10.1086/425417}, \href {https://ui.adsabs.harvard.edu/abs/2004ApJ...614L..33A} {614, L33}

\bibitem[\protect\citeauthoryear{{Arthur} et~al.,}{{Arthur} et~al.}{2019}]{Arthur2019}
{Arthur} J.,  et~al., 2019, \mn@doi [Monthly Notices of the Royal Astronomical Society] {10.1093/mnras/stz212}, \href {https://ui.adsabs.harvard.edu/abs/2019MNRAS.484.3968A} {484, 3968}

\bibitem[\protect\citeauthoryear{{Bah{\'e}} et~al.,}{{Bah{\'e}} et~al.}{2017}]{Bahe2017}
{Bah{\'e}} Y.~M.,  et~al., 2017, \mn@doi [\mnras] {10.1093/mnras/stx1403}, \href {https://ui.adsabs.harvard.edu/abs/2017MNRAS.470.4186B} {470, 4186}

\bibitem[\protect\citeauthoryear{{Benson} \& {Du}}{{Benson} \& {Du}}{2022}]{Benson2022}
{Benson} A.~J.,  {Du} X.,  2022, \mn@doi [\mnras] {10.1093/mnras/stac2750}, \href {https://ui.adsabs.harvard.edu/abs/2022MNRAS.517.1398B} {517, 1398}

\bibitem[\protect\citeauthoryear{{Boylan-Kolchin}, {Ma}  \& {Quataert}}{{Boylan-Kolchin} et~al.}{2008}]{Boylan2008}
{Boylan-Kolchin} M.,  {Ma} C.-P.,   {Quataert} E.,  2008, \mn@doi [\mnras] {10.1111/j.1365-2966.2007.12530.x}, \href {https://ui.adsabs.harvard.edu/abs/2008MNRAS.383...93B} {383, 93}

\bibitem[\protect\citeauthoryear{{Brough} et~al.,}{{Brough} et~al.}{2024}]{Brough2024}
{Brough} S.,  et~al., 2024, \mn@doi [\mnras] {10.1093/mnras/stad3810}, \href {https://ui.adsabs.harvard.edu/abs/2024MNRAS.528..771B} {528, 771}

\bibitem[\protect\citeauthoryear{Brown, Martin, Pearce, Hatch, Bahé  \& Dubois}{Brown et~al.}{2024}]{Brown2024}
Brown H.~J.,  Martin G.,  Pearce F.~R.,  Hatch N.~A.,  Bahé Y.~M.,   Dubois Y.,  2024, \mn@doi [Monthly Notices of the Royal Astronomical Society] {10.1093/mnras/stae2084}, p. stae2084

\bibitem[\protect\citeauthoryear{{Bullock}, {Dekel}, {Kolatt}, {Kravtsov}, {Klypin}, {Porciani}  \& {Primack}}{{Bullock} et~al.}{2001}]{Bullock2001}
{Bullock} J.~S.,  {Dekel} A.,  {Kolatt} T.~S.,  {Kravtsov} A.~V.,  {Klypin} A.~A.,  {Porciani} C.,   {Primack} J.~R.,  2001, \mn@doi [\apj] {10.1086/321477}, \href {https://ui.adsabs.harvard.edu/abs/2001ApJ...555..240B} {555, 240}

\bibitem[\protect\citeauthoryear{{Burke}, {Hilton}  \& {Collins}}{{Burke} et~al.}{2015}]{Burke2015}
{Burke} C.,  {Hilton} M.,   {Collins} C.,  2015, \mn@doi [\mnras] {10.1093/mnras/stv450}, \href {https://ui.adsabs.harvard.edu/abs/2015MNRAS.449.2353B} {449, 2353}

\bibitem[\protect\citeauthoryear{{Byrd} \& {Valtonen}}{{Byrd} \& {Valtonen}}{1990}]{Byrd1990}
{Byrd} G.,  {Valtonen} M.,  1990, \mn@doi [\apj] {10.1086/168362}, \href {https://ui.adsabs.harvard.edu/#abs/1990ApJ...350...89B} {350, 89}

\bibitem[\protect\citeauthoryear{{Chang}, {Macci{\`o}}  \& {Kang}}{{Chang} et~al.}{2013}]{Chang2013}
{Chang} J.,  {Macci{\`o}} A.~V.,   {Kang} X.,  2013, \mn@doi [\mnras] {10.1093/mnras/stt434}, \href {https://ui.adsabs.harvard.edu/abs/2013MNRAS.431.3533C} {431, 3533}

\bibitem[\protect\citeauthoryear{{Cheng}, {Greengard}  \& {Rokhlin}}{{Cheng} et~al.}{1999}]{Cheng1999}
{Cheng} H.,  {Greengard} L.,   {Rokhlin} V.,  1999, \mn@doi [Journal of Computational Physics] {10.1006/jcph.1999.6355}, \href {https://ui.adsabs.harvard.edu/abs/1999JCoPh.155..468C} {155, 468}

\bibitem[\protect\citeauthoryear{{Chun}, {Shin}, {Smith}, {Ko}  \& {Yoo}}{{Chun} et~al.}{2022}]{Chun2022}
{Chun} K.,  {Shin} J.,  {Smith} R.,  {Ko} J.,   {Yoo} J.,  2022, \mn@doi [\apj] {10.3847/1538-4357/ac2cbe}, \href {https://ui.adsabs.harvard.edu/abs/2022ApJ...925..103C} {925, 103}

\bibitem[\protect\citeauthoryear{{Chun}, {Shin}, {Smith}, {Ko}  \& {Yoo}}{{Chun} et~al.}{2023}]{Chun2023}
{Chun} K.,  {Shin} J.,  {Smith} R.,  {Ko} J.,   {Yoo} J.,  2023, \mn@doi [\apj] {10.3847/1538-4357/aca890}, \href {https://ui.adsabs.harvard.edu/abs/2023ApJ...943..148C} {943, 148}

\bibitem[\protect\citeauthoryear{{Chun}, {Shin}, {Ko}, {Smith}  \& {Yoo}}{{Chun} et~al.}{2024}]{Chun2024}
{Chun} K.,  {Shin} J.,  {Ko} J.,  {Smith} R.,   {Yoo} J.,  2024, \mn@doi [arXiv e-prints] {10.48550/arXiv.2405.08061}, \href {https://ui.adsabs.harvard.edu/abs/2024arXiv240508061C} {p. arXiv:2405.08061}

\bibitem[\protect\citeauthoryear{{Contini}}{{Contini}}{2021}]{Contini2021}
{Contini} E.,  2021, \mn@doi [Galaxies] {10.3390/galaxies9030060}, \href {https://ui.adsabs.harvard.edu/abs/2021Galax...9...60C} {9, 60}

\bibitem[\protect\citeauthoryear{{Contini}, {De Lucia}, {Villalobos}  \& {Borgani}}{{Contini} et~al.}{2014}]{Contini2014}
{Contini} E.,  {De Lucia} G.,  {Villalobos} {\'A}.,   {Borgani} S.,  2014, \mn@doi [\mnras] {10.1093/mnras/stt2174}, \href {https://ui.adsabs.harvard.edu/abs/2014MNRAS.437.3787C} {437, 3787}

\bibitem[\protect\citeauthoryear{{Contini}, {Kang}, {Romeo}  \& {Xia}}{{Contini} et~al.}{2017}]{Contini2017}
{Contini} E.,  {Kang} X.,  {Romeo} A.~D.,   {Xia} Q.,  2017, \mn@doi [\apj] {10.3847/1538-4357/aa5d16}, \href {https://ui.adsabs.harvard.edu/abs/2017ApJ...837...27C} {837, 27}

\bibitem[\protect\citeauthoryear{{Contini}, {Yi}  \& {Kang}}{{Contini} et~al.}{2018}]{Contini2018}
{Contini} E.,  {Yi} S.~K.,   {Kang} X.,  2018, \mn@doi [\mnras] {10.1093/mnras/sty1518}, \href {https://ui.adsabs.harvard.edu/abs/2018MNRAS.479..932C} {479, 932}

\bibitem[\protect\citeauthoryear{{Contini}, {Yi}  \& {Kang}}{{Contini} et~al.}{2019}]{Contini2019}
{Contini} E.,  {Yi} S.~K.,   {Kang} X.,  2019, \mn@doi [\apj] {10.3847/1538-4357/aaf41f}, \href {https://ui.adsabs.harvard.edu/abs/2019ApJ...871...24C} {871, 24}

\bibitem[\protect\citeauthoryear{{Contini}, {Jeon}, {Rhee}, {Han}  \& {Yi}}{{Contini} et~al.}{2023}]{Contini2023}
{Contini} E.,  {Jeon} S.,  {Rhee} J.,  {Han} S.,   {Yi} S.~K.,  2023, \mn@doi [\apj] {10.3847/1538-4357/acfd25}, \href {https://ui.adsabs.harvard.edu/abs/2023ApJ...958...72C} {958, 72}

\bibitem[\protect\citeauthoryear{{Contini}, {Rhee}, {Han}, {Jeon}  \& {Yi}}{{Contini} et~al.}{2024}]{Contini2024}
{Contini} E.,  {Rhee} J.,  {Han} S.,  {Jeon} S.,   {Yi} S.~K.,  2024, \mn@doi [\aj] {10.3847/1538-3881/ad0894}, \href {https://ui.adsabs.harvard.edu/abs/2024AJ....167....7C} {167, 7}

\bibitem[\protect\citeauthoryear{{Contreras-Santos} et~al.,}{{Contreras-Santos} et~al.}{2024}]{Contreras2024}
{Contreras-Santos} A.,  et~al., 2024, \mn@doi [arXiv e-prints] {10.48550/arXiv.2401.08283}, \href {https://ui.adsabs.harvard.edu/abs/2024arXiv240108283C} {p. arXiv:2401.08283}

\bibitem[\protect\citeauthoryear{{Cooper}, {Gao}, {Guo}, {Frenk}, {Jenkins}, {Springel}  \& {White}}{{Cooper} et~al.}{2015}]{Cooper2015}
{Cooper} A.~P.,  {Gao} L.,  {Guo} Q.,  {Frenk} C.~S.,  {Jenkins} A.,  {Springel} V.,   {White} S.~D.~M.,  2015, \mn@doi [\mnras] {10.1093/mnras/stv1042}, \href {https://ui.adsabs.harvard.edu/abs/2015MNRAS.451.2703C} {451, 2703}

\bibitem[\protect\citeauthoryear{{Crain} et~al.,}{{Crain} et~al.}{2015}]{Crain2015}
{Crain} R.~A.,  et~al., 2015, \mn@doi [\mnras] {10.1093/mnras/stv725}, \href {https://ui.adsabs.harvard.edu/abs/2015MNRAS.450.1937C} {450, 1937}

\bibitem[\protect\citeauthoryear{{Cui} et~al.,}{{Cui} et~al.}{2014}]{Cui2014}
{Cui} W.,  et~al., 2014, \mn@doi [\mnras] {10.1093/mnras/stt1940}, \href {https://ui.adsabs.harvard.edu/abs/2014MNRAS.437..816C} {437, 816}

\bibitem[\protect\citeauthoryear{{Cui} et~al.,}{{Cui} et~al.}{2018}]{Cui2018}
{Cui} W.,  et~al., 2018, \mn@doi [\mnras] {10.1093/mnras/sty2111}, \href {https://ui.adsabs.harvard.edu/abs/2018MNRAS.480.2898C} {480, 2898}

\bibitem[\protect\citeauthoryear{{Cui} et~al.,}{{Cui} et~al.}{2022}]{Cui2022}
{Cui} W.,  et~al., 2022, \mn@doi [\mnras] {10.1093/mnras/stac1402}, \href {https://ui.adsabs.harvard.edu/abs/2022MNRAS.514..977C} {514, 977}

\bibitem[\protect\citeauthoryear{{Dav{\'e}}, {Angl{\'e}s-Alc{\'a}zar}, {Narayanan}, {Li}, {Rafieferantsoa}  \& {Appleby}}{{Dav{\'e}} et~al.}{2019}]{Dave2019}
{Dav{\'e}} R.,  {Angl{\'e}s-Alc{\'a}zar} D.,  {Narayanan} D.,  {Li} Q.,  {Rafieferantsoa} M.~H.,   {Appleby} S.,  2019, \mn@doi [\mnras] {10.1093/mnras/stz937}, \href {https://ui.adsabs.harvard.edu/abs/2019MNRAS.486.2827D} {486, 2827}

\bibitem[\protect\citeauthoryear{{DeMaio}, {Gonzalez}, {Zabludoff}, {Zaritsky}, {Connor}, {Donahue}  \& {Mulchaey}}{{DeMaio} et~al.}{2018}]{DeMaio2018}
{DeMaio} T.,  {Gonzalez} A.~H.,  {Zabludoff} A.,  {Zaritsky} D.,  {Connor} T.,  {Donahue} M.,   {Mulchaey} J.~S.,  2018, \mn@doi [\mnras] {10.1093/mnras/stx2946}, \href {https://ui.adsabs.harvard.edu/abs/2018MNRAS.474.3009D} {474, 3009}

\bibitem[\protect\citeauthoryear{{Delos}}{{Delos}}{2019}]{Delos2019}
{Delos} M.~S.,  2019, \mn@doi [\prd] {10.1103/PhysRevD.100.063505}, \href {https://ui.adsabs.harvard.edu/abs/2019PhRvD.100f3505D} {100, 063505}

\bibitem[\protect\citeauthoryear{{Dubois} et~al.,}{{Dubois} et~al.}{2014}]{Dubois2014}
{Dubois} Y.,  et~al., 2014, \mn@doi [\mnras] {10.1093/mnras/stu1227}, \href {http://adsabs.harvard.edu/abs/2014MNRAS.444.1453D} {444, 1453}

\bibitem[\protect\citeauthoryear{{Dubois}, {Peirani}, {Pichon}, {Devriendt}, {Gavazzi}, {Welker}  \& {Volonteri}}{{Dubois} et~al.}{2016}]{Dubois2016}
{Dubois} Y.,  {Peirani} S.,  {Pichon} C.,  {Devriendt} J.,  {Gavazzi} R.,  {Welker} C.,   {Volonteri} M.,  2016, \mn@doi [\mnras] {10.1093/mnras/stw2265}, \href {http://adsabs.harvard.edu/abs/2016MNRAS.463.3948D} {463, 3948}

\bibitem[\protect\citeauthoryear{{Dutton} \& {Macci{\`o}}}{{Dutton} \& {Macci{\`o}}}{2014}]{Dutton2014b}
{Dutton} A.~A.,  {Macci{\`o}} A.~V.,  2014, \mn@doi [\mnras] {10.1093/mnras/stu742}, \href {https://ui.adsabs.harvard.edu/abs/2014MNRAS.441.3359D} {441, 3359}

\bibitem[\protect\citeauthoryear{{Errani} \& {Navarro}}{{Errani} \& {Navarro}}{2021}]{Errani2021}
{Errani} R.,  {Navarro} J.~F.,  2021, \mn@doi [\mnras] {10.1093/mnras/stab1215}, \href {https://ui.adsabs.harvard.edu/abs/2021MNRAS.505...18E} {505, 18}

\bibitem[\protect\citeauthoryear{{Fall} \& {Efstathiou}}{{Fall} \& {Efstathiou}}{1980}]{Fall1980}
{Fall} S.~M.,  {Efstathiou} G.,  1980, \mn@doi [\mnras] {10.1093/mnras/193.2.189}, \href {http://adsabs.harvard.edu/abs/1980MNRAS.193..189F} {193, 189}

\bibitem[\protect\citeauthoryear{{Feldmeier}, {Mihos}, {Morrison}, {Rodney}  \& {Harding}}{{Feldmeier} et~al.}{2002}]{Feldmeier2002}
{Feldmeier} J.~J.,  {Mihos} J.~C.,  {Morrison} H.~L.,  {Rodney} S.~A.,   {Harding} P.,  2002, \mn@doi [\apj] {10.1086/341472}, \href {https://ui.adsabs.harvard.edu/abs/2002ApJ...575..779F} {575, 779}

\bibitem[\protect\citeauthoryear{{Fellhauer} \& {Lin}}{{Fellhauer} \& {Lin}}{2007}]{Fellhauer2007}
{Fellhauer} M.,  {Lin} D.~N.~C.,  2007, \mn@doi [\mnras] {10.1111/j.1365-2966.2006.11308.x}, \href {https://ui.adsabs.harvard.edu/abs/2007MNRAS.375..604F} {375, 604}

\bibitem[\protect\citeauthoryear{{George} et~al.,}{{George} et~al.}{2018}]{George2018}
{George} K.,  et~al., 2018, \mn@doi [\mnras] {10.1093/mnras/sty1452}, \href {https://ui.adsabs.harvard.edu/abs/2018MNRAS.479.4126G} {479, 4126}

\bibitem[\protect\citeauthoryear{{Gnedin}}{{Gnedin}}{2003}]{Gnedin2003}
{Gnedin} O.~Y.,  2003, \mn@doi [\apj] {10.1086/344636}, \href {http://adsabs.harvard.edu/abs/2003ApJ...582..141G} {582, 141}

\bibitem[\protect\citeauthoryear{{Gnedin}, {Hernquist}  \& {Ostriker}}{{Gnedin} et~al.}{1999}]{Gnedin1999}
{Gnedin} O.~Y.,  {Hernquist} L.,   {Ostriker} J.~P.,  1999, \mn@doi [\apj] {10.1086/306910}, \href {https://ui.adsabs.harvard.edu/abs/1999ApJ...514..109G} {514, 109}

\bibitem[\protect\citeauthoryear{{Golden-Marx} et~al.,}{{Golden-Marx} et~al.}{2023}]{Marx2023}
{Golden-Marx} J.~B.,  et~al., 2023, \mn@doi [\mnras] {10.1093/mnras/stad469}, \href {https://ui.adsabs.harvard.edu/abs/2023MNRAS.521..478G} {521, 478}

\bibitem[\protect\citeauthoryear{Golden-Marx et~al.,}{Golden-Marx et~al.}{2024}]{Marx2024}
Golden-Marx J.~B.,  et~al., 2024, The Hierarchical Growth of Bright Central Galaxies and Intracluster Light as Traced by the Magnitude Gap (\mn@eprint {arXiv} {2409.02184}), \url {https://arxiv.org/abs/2409.02184}

\bibitem[\protect\citeauthoryear{{Gonzalez}, {Zaritsky}  \& {Zabludoff}}{{Gonzalez} et~al.}{2007}]{Gonzalez2007}
{Gonzalez} A.~H.,  {Zaritsky} D.,   {Zabludoff} A.~I.,  2007, \mn@doi [\apj] {10.1086/519729}, \href {https://ui.adsabs.harvard.edu/abs/2007ApJ...666..147G} {666, 147}

\bibitem[\protect\citeauthoryear{{Green}, {van den Bosch}  \& {Jiang}}{{Green} et~al.}{2021}]{Green2021}
{Green} S.~B.,  {van den Bosch} F.~C.,   {Jiang} F.,  2021, \mn@doi [\mnras] {10.1093/mnras/stab696}, \href {https://ui.adsabs.harvard.edu/abs/2021MNRAS.503.4075G} {503, 4075}

\bibitem[\protect\citeauthoryear{{Greengard} \& {Rokhlin}}{{Greengard} \& {Rokhlin}}{1987}]{Greengard1987}
{Greengard} L.,  {Rokhlin} V.,  1987, \mn@doi [Journal of Computational Physics] {10.1016/0021-9991(87)90140-9}, \href {https://ui.adsabs.harvard.edu/abs/1987JCoPh..73..325G} {73, 325}

\bibitem[\protect\citeauthoryear{{Gullieuszik} et~al.,}{{Gullieuszik} et~al.}{2020}]{Gullieuszik2020}
{Gullieuszik} M.,  et~al., 2020, \mn@doi [\apj] {10.3847/1538-4357/aba3cb}, \href {https://ui.adsabs.harvard.edu/abs/2020ApJ...899...13G} {899, 13}

\bibitem[\protect\citeauthoryear{{Guo} et~al.,}{{Guo} et~al.}{2011}]{Guo2011}
{Guo} Q.,  et~al., 2011, \mn@doi [\mnras] {10.1111/j.1365-2966.2010.18114.x}, \href {https://ui.adsabs.harvard.edu/abs/2011MNRAS.413..101G} {413, 101}

\bibitem[\protect\citeauthoryear{{Haggar}, {Pearce}, {Gray}, {Knebe}  \& {Yepes}}{{Haggar} et~al.}{2021}]{Haggar2021}
{Haggar} R.,  {Pearce} F.~R.,  {Gray} M.~E.,  {Knebe} A.,   {Yepes} G.,  2021, \mn@doi [Monthly Notices of the Royal Astronomical Society] {10.1093/mnras/stab064}, \href {https://ui.adsabs.harvard.edu/abs/2021MNRAS.502.1191H} {502, 1191}

\bibitem[\protect\citeauthoryear{{Hernquist}}{{Hernquist}}{1990}]{Hernquist1990}
{Hernquist} L.,  1990, \mn@doi [\apj] {10.1086/168845}, \href {https://ui.adsabs.harvard.edu/abs/1990ApJ...356..359H} {356, 359}

\bibitem[\protect\citeauthoryear{{Hirschmann}, {Dolag}, {Saro}, {Bachmann}, {Borgani}  \& {Burkert}}{{Hirschmann} et~al.}{2014}]{Hirschmann2014}
{Hirschmann} M.,  {Dolag} K.,  {Saro} A.,  {Bachmann} L.,  {Borgani} S.,   {Burkert} A.,  2014, \mn@doi [\mnras] {10.1093/mnras/stu1023}, \href {https://ui.adsabs.harvard.edu/abs/2014MNRAS.442.2304H} {442, 2304}

\bibitem[\protect\citeauthoryear{{Hopkins}}{{Hopkins}}{2015}]{Hopkins2015}
{Hopkins} P.~F.,  2015, \mn@doi [\mnras] {10.1093/mnras/stv195}, \href {https://ui.adsabs.harvard.edu/abs/2015MNRAS.450...53H} {450, 53}

\bibitem[\protect\citeauthoryear{{Hopkins}, {Nadler}, {Grudi{\'c}}, {Shen}, {Sands}  \& {Jiang}}{{Hopkins} et~al.}{2023}]{Hopkins2023}
{Hopkins} P.~F.,  {Nadler} E.~O.,  {Grudi{\'c}} M.~Y.,  {Shen} X.,  {Sands} I.,   {Jiang} F.,  2023, \mn@doi [\mnras] {10.1093/mnras/stad2548}, \href {https://ui.adsabs.harvard.edu/abs/2023MNRAS.525.5951H} {525, 5951}

\bibitem[\protect\citeauthoryear{{Iannuzzi} \& {Dolag}}{{Iannuzzi} \& {Dolag}}{2011}]{Iannuzzi2011}
{Iannuzzi} F.,  {Dolag} K.,  2011, \mn@doi [\mnras] {10.1111/j.1365-2966.2011.19446.x}, \href {https://ui.adsabs.harvard.edu/abs/2011MNRAS.417.2846I} {417, 2846}

\bibitem[\protect\citeauthoryear{{Jackson} et~al.,}{{Jackson} et~al.}{2024}]{Jackson2024}
{Jackson} R.~A.,  et~al., 2024, \mn@doi [\mnras] {10.1093/mnras/stae056}, \href {https://ui.adsabs.harvard.edu/abs/2024MNRAS.528.1655J} {528, 1655}

\bibitem[\protect\citeauthoryear{{Jim{\'e}nez-Teja} et~al.,}{{Jim{\'e}nez-Teja} et~al.}{2018}]{Jimenez2018}
{Jim{\'e}nez-Teja} Y.,  et~al., 2018, \mn@doi [\apj] {10.3847/1538-4357/aab70f}, \href {https://ui.adsabs.harvard.edu/abs/2018ApJ...857...79J} {857, 79}

\bibitem[\protect\citeauthoryear{{Johnston}, {Sigurdsson}  \& {Hernquist}}{{Johnston} et~al.}{1999}]{Johnston1999}
{Johnston} K.~V.,  {Sigurdsson} S.,   {Hernquist} L.,  1999, \mn@doi [\mnras] {10.1046/j.1365-8711.1999.02200.x}, \href {https://ui.adsabs.harvard.edu/abs/1999MNRAS.302..771J} {302, 771}

\bibitem[\protect\citeauthoryear{{Johnston}, {Bullock}, {Sharma}, {Font}, {Robertson}  \& {Leitner}}{{Johnston} et~al.}{2008}]{Johnston2008}
{Johnston} K.~V.,  {Bullock} J.~S.,  {Sharma} S.,  {Font} A.,  {Robertson} B.~E.,   {Leitner} S.~N.,  2008, \mn@doi [\apj] {10.1086/592228}, \href {https://ui.adsabs.harvard.edu/abs/2008ApJ...689..936J} {689, 936}

\bibitem[\protect\citeauthoryear{{Joo} \& {Jee}}{{Joo} \& {Jee}}{2023}]{Joo2023}
{Joo} H.,  {Jee} M.~J.,  2023, \mn@doi [\nat] {10.1038/s41586-022-05396-4}, \href {https://ui.adsabs.harvard.edu/abs/2023Natur.613...37J} {613, 37}

\bibitem[\protect\citeauthoryear{{Joshi}, {Parker}, {Wadsley}  \& {Keller}}{{Joshi} et~al.}{2019}]{Joshi2019}
{Joshi} G.~D.,  {Parker} L.~C.,  {Wadsley} J.,   {Keller} B.~W.,  2019, \mn@doi [\mnras] {10.1093/mnras/sty3119}, \href {https://ui.adsabs.harvard.edu/abs/2019MNRAS.483..235J} {483, 235}

\bibitem[\protect\citeauthoryear{{Kelvin} et~al.,}{{Kelvin} et~al.}{2014}]{Kelvin2014}
{Kelvin} L.~S.,  et~al., 2014, \mn@doi [\mnras] {10.1093/mnras/stu1507}, \href {https://ui.adsabs.harvard.edu/abs/2014MNRAS.444.1647K} {444, 1647}

\bibitem[\protect\citeauthoryear{{Khochfar} \& {Burkert}}{{Khochfar} \& {Burkert}}{2006}]{Khochfar2006}
{Khochfar} S.,  {Burkert} A.,  2006, \mn@doi [\aap] {10.1051/0004-6361:20053241}, \href {https://ui.adsabs.harvard.edu/abs/2006A&A...445..403K} {445, 403}

\bibitem[\protect\citeauthoryear{{Kluge} et~al.,}{{Kluge} et~al.}{2020}]{Kluge2020}
{Kluge} M.,  et~al., 2020, \mn@doi [\apjs] {10.3847/1538-4365/ab733b}, \href {https://ui.adsabs.harvard.edu/abs/2020ApJS..247...43K} {247, 43}

\bibitem[\protect\citeauthoryear{{Kluge} et~al.,}{{Kluge} et~al.}{2024}]{Kluge2024}
{Kluge} M.,  et~al., 2024, \mn@doi [arXiv e-prints] {10.48550/arXiv.2405.13503}, \href {https://ui.adsabs.harvard.edu/abs/2024arXiv240513503K} {p. arXiv:2405.13503}

\bibitem[\protect\citeauthoryear{{Knebe}, {Power}, {Gill}  \& {Gibson}}{{Knebe} et~al.}{2006}]{Knebe2006}
{Knebe} A.,  {Power} C.,  {Gill} S. P.~D.,   {Gibson} B.~K.,  2006, \mn@doi [\mnras] {10.1111/j.1365-2966.2006.10161.x}, \href {https://ui.adsabs.harvard.edu/abs/2006MNRAS.368..741K} {368, 741}

\bibitem[\protect\citeauthoryear{{Kulier} et~al.,}{{Kulier} et~al.}{2023}]{Kulier2023}
{Kulier} A.,  et~al., 2023, \mn@doi [\apj] {10.3847/1538-4357/aceda3}, \href {https://ui.adsabs.harvard.edu/abs/2023ApJ...954..177K} {954, 177}

\bibitem[\protect\citeauthoryear{{Lazar}, {Kaviraj}, {Watkins}, {Martin}, {Bichang'a}  \& {Jackson}}{{Lazar} et~al.}{2024}]{Lazar2024}
{Lazar} I.,  {Kaviraj} S.,  {Watkins} A.~E.,  {Martin} G.,  {Bichang'a} B.,   {Jackson} R.~A.,  2024, \mn@doi [\mnras] {10.1093/mnras/stae510}, \href {https://ui.adsabs.harvard.edu/abs/2024MNRAS.529..499L} {529, 499}

\bibitem[\protect\citeauthoryear{{Lynden-Bell}}{{Lynden-Bell}}{1967}]{LyndenBell1967}
{Lynden-Bell} D.,  1967, \mn@doi [\mnras] {10.1093/mnras/136.1.101}, \href {https://ui.adsabs.harvard.edu/abs/1967MNRAS.136..101L} {136, 101}

\bibitem[\protect\citeauthoryear{{Mansfield} \& {Avestruz}}{{Mansfield} \& {Avestruz}}{2021}]{Mansfield2021}
{Mansfield} P.,  {Avestruz} C.,  2021, \mn@doi [\mnras] {10.1093/mnras/staa3388}, \href {https://ui.adsabs.harvard.edu/abs/2021MNRAS.500.3309M} {500, 3309}

\bibitem[\protect\citeauthoryear{{Martin} et~al.,}{{Martin} et~al.}{2022}]{Martin2022}
{Martin} G.,  et~al., 2022, \mn@doi [\mnras] {10.1093/mnras/stac1003}, \href {https://ui.adsabs.harvard.edu/abs/2022MNRAS.513.1459M} {513, 1459}

\bibitem[\protect\citeauthoryear{{Mayer}, {Governato}  \& {Kaufmann}}{{Mayer} et~al.}{2008}]{Mayer2008}
{Mayer} L.,  {Governato} F.,   {Kaufmann} T.,  2008, \mn@doi [Advanced Science Letters] {10.48550/arXiv.0801.3845}, \href {https://ui.adsabs.harvard.edu/abs/2008ASL.....1....7M} {1, 7}

\bibitem[\protect\citeauthoryear{{McBride}, {Fakhouri}  \& {Ma}}{{McBride} et~al.}{2009}]{McBride2009}
{McBride} J.,  {Fakhouri} O.,   {Ma} C.-P.,  2009, \mn@doi [\mnras] {10.1111/j.1365-2966.2009.15329.x}, \href {https://ui.adsabs.harvard.edu/abs/2009MNRAS.398.1858M} {398, 1858}

\bibitem[\protect\citeauthoryear{{McCarthy}, {Frenk}, {Font}, {Lacey}, {Bower}, {Mitchell}, {Balogh}  \& {Theuns}}{{McCarthy} et~al.}{2008}]{McCarthy2008}
{McCarthy} I.~G.,  {Frenk} C.~S.,  {Font} A.~S.,  {Lacey} C.~G.,  {Bower} R.~G.,  {Mitchell} N.~L.,  {Balogh} M.~L.,   {Theuns} T.,  2008, \mn@doi [\mnras] {10.1111/j.1365-2966.2007.12577.x}, \href {https://ui.adsabs.harvard.edu/abs/2008MNRAS.383..593M} {383, 593}

\bibitem[\protect\citeauthoryear{{Melnick}, {Giraud}, {Toledo}, {Selman}  \& {Quintana}}{{Melnick} et~al.}{2012}]{Melnick2012}
{Melnick} J.,  {Giraud} E.,  {Toledo} I.,  {Selman} F.,   {Quintana} H.,  2012, \mn@doi [\mnras] {10.1111/j.1365-2966.2012.21924.x}, \href {https://ui.adsabs.harvard.edu/abs/2012MNRAS.427..850M} {427, 850}

\bibitem[\protect\citeauthoryear{{Mihos}}{{Mihos}}{2004}]{Mihos2004}
{Mihos} J.~C.,  2004, in {Mulchaey} J.~S.,  {Dressler} A.,   {Oemler} A.,  eds, Clusters of Galaxies: Probes of Cosmological Structure and Galaxy Evolution. p.~277

\bibitem[\protect\citeauthoryear{{Mihos}}{{Mihos}}{2016}]{Mihos2016}
{Mihos} J.~C.,  2016, in {Bragaglia} A.,  {Arnaboldi} M.,  {Rejkuba} M.,   {Romano} D.,  eds,  Proceedings of the International Astronomical Union Vol. 317, The General Assembly of Galaxy Halos: Structure, Origin and Evolution. pp 27--34 (\mn@eprint {arXiv} {1510.01929}), \mn@doi{10.1017/S1743921315006857}

\bibitem[\protect\citeauthoryear{{Mihos}, {Harding}, {Feldmeier}  \& {Morrison}}{{Mihos} et~al.}{2005}]{Mihos2005}
{Mihos} J.~C.,  {Harding} P.,  {Feldmeier} J.,   {Morrison} H.,  2005, \mn@doi [\apjl] {10.1086/497030}, \href {https://ui.adsabs.harvard.edu/abs/2005ApJ...631L..41M} {631, L41}

\bibitem[\protect\citeauthoryear{{Mihos}, {Harding}, {Feldmeier}, {Rudick}, {Janowiecki}, {Morrison}, {Slater}  \& {Watkins}}{{Mihos} et~al.}{2017}]{Mihos2017}
{Mihos} J.~C.,  {Harding} P.,  {Feldmeier} J.~J.,  {Rudick} C.,  {Janowiecki} S.,  {Morrison} H.,  {Slater} C.,   {Watkins} A.,  2017, \mn@doi [\apj] {10.3847/1538-4357/834/1/16}, \href {https://ui.adsabs.harvard.edu/abs/2017ApJ...834...16M} {834, 16}

\bibitem[\protect\citeauthoryear{{Montes}}{{Montes}}{2022}]{Montes2022}
{Montes} M.,  2022, \mn@doi [Nature Astronomy] {10.1038/s41550-022-01616-z}, \href {https://ui.adsabs.harvard.edu/abs/2022NatAs...6..308M} {6, 308}

\bibitem[\protect\citeauthoryear{{Montes} \& {Trujillo}}{{Montes} \& {Trujillo}}{2018}]{Montes2018}
{Montes} M.,  {Trujillo} I.,  2018, \mn@doi [\mnras] {10.1093/mnras/stx2847}, \href {https://ui.adsabs.harvard.edu/abs/2018MNRAS.474..917M} {474, 917}

\bibitem[\protect\citeauthoryear{{Montes}, {Brough}, {Owers}  \& {Santucci}}{{Montes} et~al.}{2021}]{Montes2021}
{Montes} M.,  {Brough} S.,  {Owers} M.~S.,   {Santucci} G.,  2021, \mn@doi [\apj] {10.3847/1538-4357/abddb6}, \href {https://ui.adsabs.harvard.edu/abs/2021ApJ...910...45M} {910, 45}

\bibitem[\protect\citeauthoryear{{Moore}, {Katz}, {Lake}, {Dressler}  \& {Oemler}}{{Moore} et~al.}{1996}]{Moore1996}
{Moore} B.,  {Katz} N.,  {Lake} G.,  {Dressler} A.,   {Oemler} A.,  1996, \mn@doi [\nat] {10.1038/379613a0}, \href {https://ui.adsabs.harvard.edu/#abs/1996Natur.379..613M} {379, 613}

\bibitem[\protect\citeauthoryear{{Morishita}, {Abramson}, {Treu}, {Schmidt}, {Vulcani}  \& {Wang}}{{Morishita} et~al.}{2017}]{Morishita2017}
{Morishita} T.,  {Abramson} L.~E.,  {Treu} T.,  {Schmidt} K.~B.,  {Vulcani} B.,   {Wang} X.,  2017, \mn@doi [\apj] {10.3847/1538-4357/aa8403}, \href {https://ui.adsabs.harvard.edu/abs/2017ApJ...846..139M} {846, 139}

\bibitem[\protect\citeauthoryear{{Moster}, {Naab}  \& {White}}{{Moster} et~al.}{2013}]{Moster2013}
{Moster} B.~P.,  {Naab} T.,   {White} S. D.~M.,  2013, \mn@doi [\mnras] {10.1093/mnras/sts261}, \href {https://ui.adsabs.harvard.edu/abs/2013MNRAS.428.3121M} {428, 3121}

\bibitem[\protect\citeauthoryear{{Murante} et~al.,}{{Murante} et~al.}{2004}]{Murante2004}
{Murante} G.,  et~al., 2004, \mn@doi [\apjl] {10.1086/421348}, \href {https://ui.adsabs.harvard.edu/abs/2004ApJ...607L..83M} {607, L83}

\bibitem[\protect\citeauthoryear{{Murante}, {Giovalli}, {Gerhard}, {Arnaboldi}, {Borgani}  \& {Dolag}}{{Murante} et~al.}{2007}]{Murante2007}
{Murante} G.,  {Giovalli} M.,  {Gerhard} O.,  {Arnaboldi} M.,  {Borgani} S.,   {Dolag} K.,  2007, \mn@doi [\mnras] {10.1111/j.1365-2966.2007.11568.x}, \href {https://ui.adsabs.harvard.edu/abs/2007MNRAS.377....2M} {377, 2}

\bibitem[\protect\citeauthoryear{{Napolitano} et~al.,}{{Napolitano} et~al.}{2003}]{Napolitano2003}
{Napolitano} N.~R.,  et~al., 2003, \mn@doi [\apj] {10.1086/376860}, \href {https://ui.adsabs.harvard.edu/abs/2003ApJ...594..172N} {594, 172}

\bibitem[\protect\citeauthoryear{{Navarro}, {Eke}  \& {Frenk}}{{Navarro} et~al.}{1996}]{Navarro1996}
{Navarro} J.~F.,  {Eke} V.~R.,   {Frenk} C.~S.,  1996, \mn@doi [\mnras] {10.1093/mnras/283.3.L72}, \href {https://ui.adsabs.harvard.edu/#abs/1996MNRAS.283L..72N} {283, L72}

\bibitem[\protect\citeauthoryear{{Nelson} et~al.,}{{Nelson} et~al.}{2019}]{Nelson2019}
{Nelson} D.,  et~al., 2019, \mn@doi [\mnras] {10.1093/mnras/stz2306}, \href {https://ui.adsabs.harvard.edu/abs/2019MNRAS.490.3234N} {490, 3234}

\bibitem[\protect\citeauthoryear{{Ogiya}, {Taylor}  \& {Hudson}}{{Ogiya} et~al.}{2021}]{Ogiya2021}
{Ogiya} G.,  {Taylor} J.~E.,   {Hudson} M.~J.,  2021, \mn@doi [\mnras] {10.1093/mnras/stab361}, \href {https://ui.adsabs.harvard.edu/abs/2021MNRAS.503.1233O} {503, 1233}

\bibitem[\protect\citeauthoryear{{Oh} et~al.,}{{Oh} et~al.}{2015}]{Oh2015}
{Oh} S.-H.,  et~al., 2015, \mn@doi [\aj] {10.1088/0004-6256/149/6/180}, \href {https://ui.adsabs.harvard.edu/abs/2015AJ....149..180O} {149, 180}

\bibitem[\protect\citeauthoryear{{Parsotan}, {Cochrane}, {Hayward}, {Angl{\'e}s-Alc{\'a}zar}, {Feldmann}, {Faucher-Gigu{\`e}re}, {Wellons}  \& {Hopkins}}{{Parsotan} et~al.}{2021}]{Parsotan2021}
{Parsotan} T.,  {Cochrane} R.~K.,  {Hayward} C.~C.,  {Angl{\'e}s-Alc{\'a}zar} D.,  {Feldmann} R.,  {Faucher-Gigu{\`e}re} C.~A.,  {Wellons} S.,   {Hopkins} P.~F.,  2021, \mn@doi [\mnras] {10.1093/mnras/staa3765}, \href {https://ui.adsabs.harvard.edu/abs/2021MNRAS.501.1591P} {501, 1591}

\bibitem[\protect\citeauthoryear{{Pe{\~n}arrubia}, {Navarro}  \& {McConnachie}}{{Pe{\~n}arrubia} et~al.}{2008}]{Penarrubia2008}
{Pe{\~n}arrubia} J.,  {Navarro} J.~F.,   {McConnachie} A.~W.,  2008, \mn@doi [\apj] {10.1086/523686}, \href {https://ui.adsabs.harvard.edu/abs/2008ApJ...673..226P} {673, 226}

\bibitem[\protect\citeauthoryear{{Pe{\~n}arrubia}, {Benson}, {Walker}, {Gilmore}, {McConnachie}  \& {Mayer}}{{Pe{\~n}arrubia} et~al.}{2010}]{Penarrubia2010}
{Pe{\~n}arrubia} J.,  {Benson} A.~J.,  {Walker} M.~G.,  {Gilmore} G.,  {McConnachie} A.~W.,   {Mayer} L.,  2010, \mn@doi [\mnras] {10.1111/j.1365-2966.2010.16762.x}, \href {https://ui.adsabs.harvard.edu/abs/2010MNRAS.406.1290P} {406, 1290}

\bibitem[\protect\citeauthoryear{{Pfeffer} \& {Baumgardt}}{{Pfeffer} \& {Baumgardt}}{2013}]{Pfeffer2013}
{Pfeffer} J.,  {Baumgardt} H.,  2013, \mn@doi [\mnras] {10.1093/mnras/stt867}, \href {https://ui.adsabs.harvard.edu/abs/2013MNRAS.433.1997P} {433, 1997}

\bibitem[\protect\citeauthoryear{{Pillepich} et~al.,}{{Pillepich} et~al.}{2014}]{Pillepich2014}
{Pillepich} A.,  et~al., 2014, \mn@doi [\mnras] {10.1093/mnras/stu1408}, \href {https://ui.adsabs.harvard.edu/abs/2014MNRAS.444..237P} {444, 237}

\bibitem[\protect\citeauthoryear{{Pillepich} et~al.,}{{Pillepich} et~al.}{2018a}]{Pillepich2018b}
{Pillepich} A.,  et~al., 2018a, \mn@doi [\mnras] {10.1093/mnras/stx2656}, \href {https://ui.adsabs.harvard.edu/abs/2018MNRAS.473.4077P} {473, 4077}

\bibitem[\protect\citeauthoryear{{Pillepich} et~al.,}{{Pillepich} et~al.}{2018b}]{Pillepich2018}
{Pillepich} A.,  et~al., 2018b, \mn@doi [\mnras] {10.1093/mnras/stx3112}, \href {https://ui.adsabs.harvard.edu/abs/2018MNRAS.475..648P} {475, 648}

\bibitem[\protect\citeauthoryear{{Power}, {Navarro}, {Jenkins}, {Frenk}, {White}, {Springel}, {Stadel}  \& {Quinn}}{{Power} et~al.}{2003}]{Power2003}
{Power} C.,  {Navarro} J.~F.,  {Jenkins} A.,  {Frenk} C.~S.,  {White} S.~D.~M.,  {Springel} V.,  {Stadel} J.,   {Quinn} T.,  2003, \mn@doi [\mnras] {10.1046/j.1365-8711.2003.05925.x}, \href {https://ui.adsabs.harvard.edu/abs/2003MNRAS.338...14P} {338, 14}

\bibitem[\protect\citeauthoryear{{Press} \& {Schechter}}{{Press} \& {Schechter}}{1974}]{Press1974}
{Press} W.~H.,  {Schechter} P.,  1974, \mn@doi [\apj] {10.1086/152650}, \href {https://ui.adsabs.harvard.edu/abs/1974ApJ...187..425P} {187, 425}

\bibitem[\protect\citeauthoryear{{Price} \& {Monaghan}}{{Price} \& {Monaghan}}{2007}]{Price2007}
{Price} D.~J.,  {Monaghan} J.~J.,  2007, \mn@doi [\mnras] {10.1111/j.1365-2966.2006.11241.x}, \href {https://ui.adsabs.harvard.edu/abs/2007MNRAS.374.1347P} {374, 1347}

\bibitem[\protect\citeauthoryear{{Proctor}, {Lagos}, {Ludlow}  \& {Robotham}}{{Proctor} et~al.}{2024}]{Proctor2024}
{Proctor} K.~L.,  {Lagos} C. d.~P.,  {Ludlow} A.~D.,   {Robotham} A. S.~G.,  2024, \mn@doi [\mnras] {10.1093/mnras/stad3341}, \href {https://ui.adsabs.harvard.edu/abs/2024MNRAS.527.2624P} {527, 2624}

\bibitem[\protect\citeauthoryear{{Puchwein}, {Springel}, {Sijacki}  \& {Dolag}}{{Puchwein} et~al.}{2010}]{Puchwein2010}
{Puchwein} E.,  {Springel} V.,  {Sijacki} D.,   {Dolag} K.,  2010, \mn@doi [\mnras] {10.1111/j.1365-2966.2010.16786.x}, \href {https://ui.adsabs.harvard.edu/abs/2010MNRAS.406..936P} {406, 936}

\bibitem[\protect\citeauthoryear{{Purcell}, {Bullock}  \& {Zentner}}{{Purcell} et~al.}{2007}]{Purcell2007}
{Purcell} C.~W.,  {Bullock} J.~S.,   {Zentner} A.~R.,  2007, \mn@doi [\apj] {10.1086/519787}, \href {https://ui.adsabs.harvard.edu/abs/2007ApJ...666...20P} {666, 20}

\bibitem[\protect\citeauthoryear{{Ragusa} et~al.,}{{Ragusa} et~al.}{2023}]{Ragusa2023}
{Ragusa} R.,  et~al., 2023, \mn@doi [\aap] {10.1051/0004-6361/202245530}, \href {https://ui.adsabs.harvard.edu/abs/2023A&A...670L..20R} {670, L20}

\bibitem[\protect\citeauthoryear{{Rudick}, {Mihos}  \& {McBride}}{{Rudick} et~al.}{2006}]{Rudick2006}
{Rudick} C.~S.,  {Mihos} J.~C.,   {McBride} C.,  2006, \mn@doi [\apj] {10.1086/506176}, \href {https://ui.adsabs.harvard.edu/abs/2006ApJ...648..936R} {648, 936}

\bibitem[\protect\citeauthoryear{{Rudick}, {Mihos}, {Harding}, {Feldmeier}, {Janowiecki}  \& {Morrison}}{{Rudick} et~al.}{2010}]{Rudick2010}
{Rudick} C.~S.,  {Mihos} J.~C.,  {Harding} P.,  {Feldmeier} J.~J.,  {Janowiecki} S.,   {Morrison} H.~L.,  2010, \mn@doi [\apj] {10.1088/0004-637X/720/1/569}, \href {https://ui.adsabs.harvard.edu/abs/2010ApJ...720..569R} {720, 569}

\bibitem[\protect\citeauthoryear{{Rudick}, {Mihos}  \& {McBride}}{{Rudick} et~al.}{2011}]{Rudick2011}
{Rudick} C.~S.,  {Mihos} J.~C.,   {McBride} C.~K.,  2011, \mn@doi [\apj] {10.1088/0004-637X/732/1/48}, \href {https://ui.adsabs.harvard.edu/abs/2011ApJ...732...48R} {732, 48}

\bibitem[\protect\citeauthoryear{{Schaller}, {Gonnet}, {Draper}, {Chalk}, {Bower}, {Willis}  \& {Hausammann}}{{Schaller} et~al.}{2018}]{Schaller2018}
{Schaller} M.,  {Gonnet} P.,  {Draper} P.~W.,  {Chalk} A. B.~G.,  {Bower} R.~G.,  {Willis} J.,   {Hausammann} L.,  2018, {SWIFT: SPH With Inter-dependent Fine-grained Tasking}, Astrophysics Source Code Library, record ascl:1805.020

\bibitem[\protect\citeauthoryear{{Schaller} et~al.,}{{Schaller} et~al.}{2023}]{Schaller2023}
{Schaller} M.,  et~al., 2023, \mn@doi [arXiv e-prints] {10.48550/arXiv.2305.13380}, \href {https://ui.adsabs.harvard.edu/abs/2023arXiv230513380S} {p. arXiv:2305.13380}

\bibitem[\protect\citeauthoryear{{Schaye} et~al.,}{{Schaye} et~al.}{2015}]{Schaye2015}
{Schaye} J.,  et~al., 2015, \mn@doi [\mnras] {10.1093/mnras/stu2058}, \href {http://adsabs.harvard.edu/abs/2015MNRAS.446..521S} {446, 521}

\bibitem[\protect\citeauthoryear{{Schechter}}{{Schechter}}{1976}]{Schechter1976}
{Schechter} P.,  1976, \mn@doi [\apj] {10.1086/154079}, \href {https://ui.adsabs.harvard.edu/#abs/1976ApJ...203..297S} {203, 297}

\bibitem[\protect\citeauthoryear{{Sedgwick}, {Baldry}, {James}  \& {Kelvin}}{{Sedgwick} et~al.}{2019}]{Sedgwick2019}
{Sedgwick} T.~M.,  {Baldry} I.~K.,  {James} P.~A.,   {Kelvin} L.~S.,  2019, \mn@doi [\mnras] {10.1093/mnras/stz186}, \href {https://ui.adsabs.harvard.edu/abs/2019MNRAS.484.5278S} {484, 5278}

\bibitem[\protect\citeauthoryear{{Smith} et~al.,}{{Smith} et~al.}{2015}]{Smith2015}
{Smith} R.,  et~al., 2015, \mn@doi [\mnras] {10.1093/mnras/stv2082}, \href {https://ui.adsabs.harvard.edu/abs/2015MNRAS.454.2502S} {454, 2502}

\bibitem[\protect\citeauthoryear{{Smith}, {Choi}, {Lee}, {Rhee}, {Sanchez-Janssen}  \& {Yi}}{{Smith} et~al.}{2016}]{Smith2016}
{Smith} R.,  {Choi} H.,  {Lee} J.,  {Rhee} J.,  {Sanchez-Janssen} R.,   {Yi} S.~K.,  2016, \mn@doi [\apj] {10.3847/1538-4357/833/1/109}, \href {https://ui.adsabs.harvard.edu/abs/2016ApJ...833..109S} {833, 109}

\bibitem[\protect\citeauthoryear{{Sommer-Larsen}, {Romeo}  \& {Portinari}}{{Sommer-Larsen} et~al.}{2005}]{Sommer2005}
{Sommer-Larsen} J.,  {Romeo} A.~D.,   {Portinari} L.,  2005, \mn@doi [\mnras] {10.1111/j.1365-2966.2005.08599.x}, \href {https://ui.adsabs.harvard.edu/abs/2005MNRAS.357..478S} {357, 478}

\bibitem[\protect\citeauthoryear{{Spitzer}}{{Spitzer}}{1942}]{Spitzer1942}
{Spitzer} Lyman J.,  1942, \mn@doi [\apj] {10.1086/144407}, \href {https://ui.adsabs.harvard.edu/abs/1942ApJ....95..329S} {95, 329}

\bibitem[\protect\citeauthoryear{{Springel} \& {White}}{{Springel} \& {White}}{1999}]{Springel1999}
{Springel} V.,  {White} S. D.~M.,  1999, \mn@doi [\mnras] {10.1046/j.1365-8711.1999.02613.x}, \href {https://ui.adsabs.harvard.edu/abs/1999MNRAS.307..162S} {307, 162}

\bibitem[\protect\citeauthoryear{{Springel} et~al.,}{{Springel} et~al.}{2008}]{Springel2008}
{Springel} V.,  et~al., 2008, \mn@doi [\mnras] {10.1111/j.1365-2966.2008.14066.x}, \href {https://ui.adsabs.harvard.edu/abs/2008MNRAS.391.1685S} {391, 1685}

\bibitem[\protect\citeauthoryear{{St{\"u}cker}, {Ogiya}, {Angulo}, {Aguirre-Santaella}  \& {S{\'a}nchez-Conde}}{{St{\"u}cker} et~al.}{2023}]{Stucker2023}
{St{\"u}cker} J.,  {Ogiya} G.,  {Angulo} R.~E.,  {Aguirre-Santaella} A.,   {S{\'a}nchez-Conde} M.~A.,  2023, \mn@doi [\mnras] {10.1093/mnras/stad844}, \href {https://ui.adsabs.harvard.edu/abs/2023MNRAS.521.4432S} {521, 4432}

\bibitem[\protect\citeauthoryear{{Tang}, {Lin}, {Cui}, {Kang}, {Wang}, {Contini}  \& {Yu}}{{Tang} et~al.}{2018}]{Tang2018}
{Tang} L.,  {Lin} W.,  {Cui} W.,  {Kang} X.,  {Wang} Y.,  {Contini} E.,   {Yu} Y.,  2018, \mn@doi [\apj] {10.3847/1538-4357/aabd78}, \href {https://ui.adsabs.harvard.edu/abs/2018ApJ...859...85T} {859, 85}

\bibitem[\protect\citeauthoryear{{Teyssier}}{{Teyssier}}{2002}]{Teyssier2002}
{Teyssier} R.,  2002, \mn@doi [\aap] {10.1051/0004-6361:20011817}, \href {http://adsabs.harvard.edu/abs/2002A%26A...385..337T} {385, 337}

\bibitem[\protect\citeauthoryear{{Tollet}, {Cattaneo}, {Mamon}, {Moutard}  \& {van den Bosch}}{{Tollet} et~al.}{2017}]{Tollet2017}
{Tollet} {\'E}.,  {Cattaneo} A.,  {Mamon} G.~A.,  {Moutard} T.,   {van den Bosch} F.~C.,  2017, \mn@doi [\mnras] {10.1093/mnras/stx1840}, \href {https://ui.adsabs.harvard.edu/abs/2017MNRAS.471.4170T} {471, 4170}

\bibitem[\protect\citeauthoryear{{Tonnesen} \& {Bryan}}{{Tonnesen} \& {Bryan}}{2012}]{Tonnesen2012}
{Tonnesen} S.,  {Bryan} G.~L.,  2012, \mn@doi [\mnras] {10.1111/j.1365-2966.2012.20737.x}, \href {https://ui.adsabs.harvard.edu/abs/2012MNRAS.422.1609T} {422, 1609}

\bibitem[\protect\citeauthoryear{{Vacondio}, {Rogers}, {Stansby}, {Mignosa}  \& {Feldman}}{{Vacondio} et~al.}{2013}]{Vacondio2013}
{Vacondio} R.,  {Rogers} B.~D.,  {Stansby} P.~K.,  {Mignosa} P.,   {Feldman} J.,  2013, \mn@doi [Computer Methods in Applied Mechanics and Engineering] {10.1016/j.cma.2012.12.014}, \href {https://ui.adsabs.harvard.edu/abs/2013CMAME.256..132V} {256, 132}

\bibitem[\protect\citeauthoryear{{Vilchez-Gomez}, {Pello}  \& {Sanahuja}}{{Vilchez-Gomez} et~al.}{1994}]{Gomez1994}
{Vilchez-Gomez} R.,  {Pello} R.,   {Sanahuja} B.,  1994, \aap, \href {https://ui.adsabs.harvard.edu/abs/1994A&A...283...37V} {283, 37}

\bibitem[\protect\citeauthoryear{{Villalobos}, {De Lucia}, {Borgani}  \& {Murante}}{{Villalobos} et~al.}{2012}]{Villalobos2012}
{Villalobos} {\'A}.,  {De Lucia} G.,  {Borgani} S.,   {Murante} G.,  2012, \mn@doi [\mnras] {10.1111/j.1365-2966.2012.20667.x}, \href {https://ui.adsabs.harvard.edu/abs/2012MNRAS.424.2401V} {424, 2401}

\bibitem[\protect\citeauthoryear{{Vulcani} et~al.,}{{Vulcani} et~al.}{2013}]{Vulcani2013}
{Vulcani} B.,  et~al., 2013, \mn@doi [\aap] {10.1051/0004-6361/201118388}, \href {https://ui.adsabs.harvard.edu/abs/2013A&A...550A..58V} {550, A58}

\bibitem[\protect\citeauthoryear{{Watson}, {Berlind}  \& {Zentner}}{{Watson} et~al.}{2012}]{Watson2012}
{Watson} D.~F.,  {Berlind} A.~A.,   {Zentner} A.~R.,  2012, \mn@doi [\apj] {10.1088/0004-637X/754/2/90}, \href {https://ui.adsabs.harvard.edu/abs/2012ApJ...754...90W} {754, 90}

\bibitem[\protect\citeauthoryear{Wendland}{Wendland}{1995}]{Wendland1995}
Wendland H.,  1995, Advances in computational Mathematics, 4, 389

\bibitem[\protect\citeauthoryear{{Wetzel}}{{Wetzel}}{2011}]{Wetzel2011}
{Wetzel} A.~R.,  2011, \mn@doi [\mnras] {10.1111/j.1365-2966.2010.17877.x}, \href {https://ui.adsabs.harvard.edu/abs/2011MNRAS.412...49W} {412, 49}

\bibitem[\protect\citeauthoryear{{Willman}, {Governato}, {Wadsley}  \& {Quinn}}{{Willman} et~al.}{2004}]{Willman2004}
{Willman} B.,  {Governato} F.,  {Wadsley} J.,   {Quinn} T.,  2004, \mn@doi [\mnras] {10.1111/j.1365-2966.2004.08312.x}, \href {https://ui.adsabs.harvard.edu/abs/2004MNRAS.355..159W} {355, 159}

\bibitem[\protect\citeauthoryear{{Yoo}, {Ko}, {Kim}  \& {Kim}}{{Yoo} et~al.}{2021}]{Yoo2021}
{Yoo} J.,  {Ko} J.,  {Kim} J.-W.,   {Kim} H.,  2021, \mn@doi [\mnras] {10.1093/mnras/stab2707}, \href {https://ui.adsabs.harvard.edu/abs/2021MNRAS.508.2634Y} {508, 2634}

\bibitem[\protect\citeauthoryear{{Yoo} et~al.,}{{Yoo} et~al.}{2024}]{Yoo2024}
{Yoo} J.,  et~al., 2024, \mn@doi [arXiv e-prints] {10.48550/arXiv.2402.17958}, \href {https://ui.adsabs.harvard.edu/abs/2024arXiv240217958Y} {p. arXiv:2402.17958}

\bibitem[\protect\citeauthoryear{{Yurin} \& {Springel}}{{Yurin} \& {Springel}}{2014}]{Yurin2014}
{Yurin} D.,  {Springel} V.,  2014, \mn@doi [\mnras] {10.1093/mnras/stu1421}, \href {https://ui.adsabs.harvard.edu/abs/2014MNRAS.444...62Y} {444, 62}

\bibitem[\protect\citeauthoryear{{Zhang} et~al.,}{{Zhang} et~al.}{2019}]{Zhang2019}
{Zhang} Y.,  et~al., 2019, \mn@doi [\apj] {10.3847/1538-4357/ab0dfd}, \href {https://ui.adsabs.harvard.edu/abs/2019ApJ...874..165Z} {874, 165}

\bibitem[\protect\citeauthoryear{{Zibetti}, {White}, {Schneider}  \& {Brinkmann}}{{Zibetti} et~al.}{2005}]{Zibetti2005}
{Zibetti} S.,  {White} S. D.~M.,  {Schneider} D.~P.,   {Brinkmann} J.,  2005, \mn@doi [\mnras] {10.1111/j.1365-2966.2005.08817.x}, \href {https://ui.adsabs.harvard.edu/abs/2005MNRAS.358..949Z} {358, 949}

\bibitem[\protect\citeauthoryear{{van Albada}}{{van Albada}}{1982}]{Albada1982}
{van Albada} T.~S.,  1982, \mn@doi [\mnras] {10.1093/mnras/201.4.939}, \href {https://ui.adsabs.harvard.edu/abs/1982MNRAS.201..939V} {201, 939}

\bibitem[\protect\citeauthoryear{{van Kampen}}{{van Kampen}}{2000}]{vanKampen2000}
{van Kampen} E.,  2000, \mn@doi [arXiv e-prints] {10.48550/arXiv.astro-ph/0002027}, \href {https://ui.adsabs.harvard.edu/abs/2000astro.ph..2027V} {pp astro--ph/0002027}

\bibitem[\protect\citeauthoryear{{van den Bosch} \& {Ogiya}}{{van den Bosch} \& {Ogiya}}{2018}]{vandenBosch2018}
{van den Bosch} F.~C.,  {Ogiya} G.,  2018, \mn@doi [\mnras] {10.1093/mnras/sty084}, \href {https://ui.adsabs.harvard.edu/abs/2018MNRAS.475.4066V} {475, 4066}

\bibitem[\protect\citeauthoryear{{van der Burg}, {McGee}, {Aussel}, {Dahle}, {Arnaud}, {Pratt}  \& {Muzzin}}{{van der Burg} et~al.}{2018}]{vanderBurg2018}
{van der Burg} R. F.~J.,  {McGee} S.,  {Aussel} H.,  {Dahle} H.,  {Arnaud} M.,  {Pratt} G.~W.,   {Muzzin} A.,  2018, \mn@doi [\aap] {10.1051/0004-6361/201833572}, \href {https://ui.adsabs.harvard.edu/abs/2018A&A...618A.140V} {618, A140}

\makeatother
\end{thebibliography}

%%%%%%%%%%%%%%%%%%%%%%%%%%%%%%%%%%%%%%%%%%%%%%%%%%

%%%%%%%%%%%%%%%%% APPENDICES %%%%%%%%%%%%%%%%%%%%%

%\appendix

%\section{Low-mass slope}
%\label{appendix:slope}

%\begin{figure}
%    \centering
%    \includegraphics[width=0.45\textwidth]{Figures/slope.pdf}
%    \caption{Fraction of total ICL mass recovered compared with the benchmark run for different resolution runs assuming GSMFs with a range of low-mass slopes. Our fiducial choice of $\alpha=-1.4$ is marked by an open square symbol.}
    %\label{fig:slope}
%\end{figure}


%%%%%%%%%%%%%%%%%%%%%%%%%%%%%%%%%%%%%%%%%%%%%%%%%%

% Don't change these lines
\bsp	% typesetting comment
\label{lastpage}
\end{document}